\documentclass[a4paper,11pt,nofootinbib]{revtex4}

\usepackage{amsfonts}
\usepackage{amssymb}
\usepackage{amsmath}
\usepackage{bbm}

\newcommand*{\dd}{\mathrm{d}}
\newcommand*{\prt}{\partial}
\newcommand*{\be}{\begin{equation}}
\newcommand*{\ee}{\end{equation}}
\newcommand*{\bea}{\begin{eqnarray}}
\newcommand*{\eea}{\end{eqnarray}}
\newcommand*{\pt}{\ \mathrm{.}}
\newcommand*{\pc}{\ \mathrm{,}}

\newcommand*{\mand}{\quad \mathrm{and} \quad}

\newcommand*{\mt}{\mathrm}
\newcommand*{\tr}{\mathrm{tr}}
\newcommand*{\Tr}{\mathrm{Tr}}

\newcommand*{\sgn}{\mathrm{sgn}}

\newcommand*{\1}{\mathbbm{1}}
\newcommand*{\lo}{l_{\circ}}

\begin{document}

\title{Quantum mechanics on $SO(3)$ via non-commutative dual variables}

\author{Daniele Oriti}
\email{daniele.oriti@aei.mpg.de}
\author{Matti Raasakka}
\email{matti.raasakka@aei.mpg.de}
\affiliation{Max Planck Institute for Gravitational Physics (Albert Einstein Institute), Am M\"uhlenberg 1, D-14476 Golm, Germany, EU}

\date{\today}

\begin{abstract}
We formulate quantum mechanics on the group $SO(3)$ using a non-commutative dual space representation for the quantum states, inspired by recent work in quantum gravity. The new non-commutative variables have a clear connection to the corresponding classical variables, and our analysis confirms them as the natural phase space variables, both mathematically and physically. In particular,  we derive the first order (Hamiltonian) path integral in terms of the non-commutative variables, as a formulation of the transition amplitudes alternative to that based on harmonic analysis. We find that the non-trivial phase space structure gives naturally rise to quantum corrections to the action for which we find a closed expression. We then study both the semi-classical approximation of the first order path integral and the example of a free particle on $SO(3)$. On the basis of these results, we comment on the relevance of similar structures and methods for more complicated theories with group-based configuration spaces, such as Loop Quantum Gravity and Spin Foam models.
\end{abstract}

\preprint{AEI-2011-013}

\maketitle

\section{Introduction}
Quantum mechanics on a Lie group has been considered, following the seminal work of DeWitt \cite{DeWitt} on quantum mechanics on general curved manifolds, first by Schulman \cite{Schulman} for $SU(2)$ and later by Dowker \cite{Dowker} for semi-simple Lie groups. This work has been later expanded upon, e.g., by Marinov and Terentyev \cite{MarinovTerentyev}. In the case of group manifolds, the group structure, and thus ensuing homogeneity and representation theory, admit a considerable simplification compared to the general case considered by DeWitt. These formulations are considered largely satisfactory, apart from some disagreement about quantum correction terms in the path integral formulation \cite{DeWitt,McLaughlinSchulman,KleinertShabanov}. We will show that our approach produces quantum corrections consistent with those obtained originally by DeWitt \cite{DeWitt}, and find that this form is required for the propagator to satisfy the Schr\"odinger equation.

Moreover, complications in considering quantum systems on Lie groups are bound to arise also from the rather involved representation theory of Lie groups, which is used in the spectral decomposition of quantum states via Peter-Weyl theorem. Recently, a new way of decomposing fields on $SO(3)$ into unitary representations of the group, the so-called group Fourier transform, was introduced in the context of 3d quantum gravity models in \cite{FreidelLivine2,FreidelLivine}, further developed in \cite{FreidelMajid}, and generalized to $SU(2)$ by Joung, Mourad and Noui \cite{JoungMouradNoui}. The group Fourier transform, an isometry from the $L^{2}$-space of functions on $SO(3)$ to an $L^{2}$-space of functions on $\mathbb{R}^{3}$ equipped with a non-commutative $\star$-product structure (reviewed in Section \ref{sec:groupfourier}), is based on the theory of quantum groups (Hopf algebras) and the corresponding definition of non-commutative geometry \cite{majid}. It has the convenient property of circumventing the representation theory by encoding the non-Abelian group structure in the non-commutative structure of the dual space. The transform is yet to be formulated for a general Lie group, but this appears as merely a question of further research, not a fundamental difficulty. Therefore, it is particularly interesting to apply the group Fourier transform technique to well-known systems, such as quantum mechanics on $SO(3)$, to be able to compare the results, and to gain more insight into the interpretation of the new non-commutative variables. This is the main purpose of this work.

Our more general motivation comes, however, from quantum gravity. A special case of quantum mechanics on a Lie group, even though it is made rather unusual by background independence, is Loop Quantum Gravity (LQG), which is the theory obtained via canonical quantization of the general theory of relativity \cite{Thiemann,carlo}. In LQG the kinematical Hilbert space of quantum states is given, in fact, in terms of cylindrical functions associated to graphs (embedded in 3-space) on a configuration space $SO(3)^{L}/SO(3)^{V}$, where $L$ is the number of links of the graph, and $V$ the number of vertices. By harmonic analysis, the same states can be recast in the form of spin networks, i.e., graphs labelled by irreducible representations of $SO(3)$. Analogously, an approach to the dynamics of LQG uses 2-complexes (as a spacetime counterpart of the graphs on which states are defined) to represent a discrete substitute for continuum spacetime, decorated either by group elements (as in lattice gauge theory) or by representations. These labelled 2-complexes are called \textit{spin foams}. One then defines quantum amplitudes for these spin foams to give a purely algebraic version of a gravitational path integral \cite{SF}. The group Fourier transform has recently been applied both to Loop Quantum Gravity states \cite{NCflux}, to define a representation of the states in terms of non-commutative fluxes of the gravitational triad field \cite{Thiemann}, and to the spin foam dynamics \cite{BaratinOriti,GFTdiffeos} or, more specifically, to the more general context of Group Field Theory \cite{iogft,ProcCapeTown}, where the transform has been shown to encode an exact duality between Spin Foam models and lattice gravity path integrals, thus making the geometry of the spin foam dynamics manifest. 
We will give some more details on these quantum gravity constructions, and comment on the relation and implications of our considerations to LQG and Spin Foam models along the way.

The layout of the paper is following: In Section \ref{sec:class} we will review the canonical formulation of classical mechanics on $SO(3)$, on which we build up our treatment of quantum mechanics on $SO(3)$ in Section \ref{sec:qm}. We start from the canonical quantization, first in terms of group variables in Subsection \ref{sec:groupbasis}, in terms of spin labels in Subsection \ref{sec:spins}, and finally in the dual non-commutative space in Subsection \ref{sec:momentumbasis} obtained via the group Fourier transform reviewed in Subsection \ref{sec:groupfourier}. In Subsection \ref{sec:pathintegral} we then derive the first order phase space path integral in terms of the group and dual non-commutative variables, and show in Subsection \ref{sec:semiclassicalanalysis} that it yields the correct semi-classical behavior. In Subsection \ref{sec:freeparticle} we further demonstrate that in the well-known case of a free quantum particle on $SO(3)$ the amplitudes coincide with the previous results in the literature with the correct form for the quantum corrections to the action. Section \ref{sec:conclusions} wraps up with the conclusions from our results, as well as some insights they provide on quantum gravity.

\section{Classical mechanics on $SO(3)$}\label{sec:class}
Classical mechanics of a non-relativistic physical system, evolving in a global time $t$, and whose configuration space is $SO(3)$, can be rigorously formulated by starting from the cotangent bundle of $SO(3)$ (the phase space) and its canonical symplectic structure. We will shortly review this construction in this section. More details can be found, e.g., in \cite{Nakahara,MarsdenRatiu,Vilasi}.

\subsection{Structure of the phase space $\mathcal{T}^{*}SO(3)$}
As a Lie group, $SO(3)$ has the convenient property of having a parallelizable cotangent bundle $\mathcal{T}^{*}SO(3) \cong SO(3) \times \mathfrak{so}(3)$ due to the existence of left-(or right-)invariant 1-forms $P_{g}$ induced by the group action as
\be
	P_{gh} \equiv L^{*}_{g}P_{h} \in \mathcal{T}^{*}_{gh}SO(3) \pc
\ee
where $L^{*}$ is the pull-back of the left multiplication $L_{g}h \equiv gh$ on 1-forms. Thus, the cotangent bundle is parallelized by the left-invariant 1-forms labelled by covectors $P_{e} =: P$ in the cotangent space at the unit element $\mathcal{T}^{*}_{e}SO(3) \cong \mathfrak{so}(3) \cong \mathbb{R}^{3}$, which we accordingly interpret as the momentum space of the system.\footnote{In fact, choosing a specific polarization of the phase space, namely, the physical interpretation of which part of it corresponds to the configuration space and which part to the momentum space, is not crucial at this point. Indeed, there is a mathematical duality between the two parts, namely, one part corresponds to the generators of translations of the other via the canonical symplectic structure of the cotangent bundle (to be introduced in the following). The physical meaning to the cotangent bundle is ultimately attached by the dynamics, e.g., the Hamiltonian of the physical system in consideration and, in particular, its symmetries. Typically, the physical configuration space is considered to be the part of the phase space in which the dynamics of a free system is translationally invariant. Nevertheless, in what follows we will use, for the sake of simplicity, the labels `configuration space' and `momentum space' for the parts $SO(3)$ and $\mathfrak{so}(3)$ of the phase space $SO(3) \times \mathfrak{so}(3)$, respectively, bearing in mind that the labelling is always invertible when no particular physical interpretation is given. Indeed, notice that in the papers \cite{FreidelLivine,FreidelMajid,JoungMouradNoui} from which our treament of the non-commutative dual variables stems, the opposite polarization has been considered, due to the specific form of the action of the field theory considered.}

A metric 2-form in the tangent bundle $\mathcal{T}SO(3)$ is obtained as the pull-back with respect to left multiplication of the Killing form $B(X,Y) := (\lo^{2}/2) \tr(XY)$, $X,Y \in \mathfrak{so}(3)$, from $\mathcal{T}_{e}SO(3) \cong \mathfrak{so}(3)$ onto the whole tangent bundle.\footnote{In this paper we use the physics convention of Lie algebras, where $\mathfrak{so}(3)$ consists of traceless Hermitean elements.} Here, $\tr(XY)$ is taken in the fundamental representation so that the left-invariant vector fields given by $\lo^{-1} T_{i,g} \equiv \lo^{-1} L_{*g} T_{i,e}$, where $t_{i} := T_{i,e} = \sigma_{i}$ in the fundamental representation, $\sigma_{i}$ being the Pauli matrices, constitute an orthonormal set of vector fields in $\mathcal{T}SO(3)$. Similarly, on the cotangent bundle $\mathcal{T}^{*}SO(3)$ a metric tensor is obtained as the push-forward with respect to left multiplication of the form $\tilde{B}(P,Q) := (\lo^{-2}/2) \tr(PQ)$ in $\mathcal{T}_{e}^{*}SO(3) \cong \mathfrak{so}(3)$, and an orthonormal basis of left-invariant 1-forms is given by $\lo T^{i}_{g} \equiv \lo L^{*}_{g} T^{i}_{e}$, where $t^{i} := T^{i}_{e} = \sigma_{i}$. $\lo \in \mathbb{R}_{+}$ is a constant with dimensions of length, which determines the length scale on the group manifold, namely, $\lo$ is the radius of the 3-sphere $S^{3} \cong SU(2) \cong SO(3) \times \mathbb{Z}_{2}$ embedded into a 4-dimensional Euclidean space, to which $SO(3)$ corresponds when antipodes $g,-g \in SU(2)$ are identified. Then, consistency requires us to choose the integration measure $\pi^{2}\lo^{3}\dd g$ on $SO(3)$, where $\dd g$ is the left-invariant Haar measure normalized to unity, so that the volume of the group, $\pi^{2}\lo^{3}$, is the volume of the upper hemisphere of $S^{3}$ with radius $\lo$. 

The Lie brackets of the left-invariant orthonormal basis vector fields $\lo^{-1}T_{i}$ reflect the structure of the underlying $\mathfrak{so}(3)$ Lie algebra:
\be
	[\mathcal{L}_{\lo^{-1}T_{i}},\mathcal{L}_{\lo^{-1}T_{j}}] = -2\lo^{-1}\epsilon_{ij}^{\phantom{ij}k}\mathcal{L}_{\lo^{-1}T_{k}} \pc
\ee
where $\mathcal{L}_{T_{i}}f(g) := \left. \frac{\dd}{\dd s}f(ge^{ist_{i}})\right|_{s=0}$ is the Lie derivative of a function $f \in C^{1}(SO(3))$ with respect to the left-invariant vector field $T_{i}$. We have here chosen the fields $T_{i}$ to be dimensionless while the dimensions are given by the constant $\lo$. By first absorbing the constant into the orthonormal basis $\lo^{-1} T_{i} \mapsto T_{i}$ and into the definition of trace $\lo^{2}\tr \mapsto \tr$ in the Killing metric, we may take the Euclidean limit $\lo \rightarrow \infty$, where Lie derivatives become commutative and the integration measure becomes the Lebesgue measure on $\mathbb{R}^{3}$, while the basis fields remain orthonormal. Thus, in this limit $SO(3)$ becomes essentially Euclidean $\mathbb{R}^{3}$ (even though the global topology of the manifold does not change\footnote{We thank Frank Eckert for pointing this out.}).

On the phase space we have the canonical 1-form $\theta \in \mathcal{T}^{*}(\mathcal{T}^{*}SO(3)) \cong \mathcal{T}^{*}SO(3) \times \mathcal{T}SO(3)$ given by $\theta_{P_{g}} \equiv (P_{g}, 0) \in \mathcal{T}^{*}_{P_{g}}(\mathcal{T}^{*}SO(3))$ for all $P_{g} \in \mathcal{T}^{*}SO(3)$, where $0$ is the null vector field on $SO(3)$. The symplectic 2-form $\omega$ on the phase space is given via the exterior derivative of the canonical 1-form $\theta$ as
\be\label{eq:symp2-form}
	\omega := -\dd\theta = \dd g^{i} \wedge \dd P_{i} - 2\lo^{-1}\epsilon_{ij}^{\phantom{ij}k}P_{k} \dd g^{i} \wedge \dd g^{j} \pc
\ee
where we used the notation $\dd g^{i} := (\lo T^{i},0)$, $\dd P_{i} := (0,\lo^{-1}T_{i})$.\footnote{The Darboux theorem ensures the existence of a coordinate system $(x^{i},p_{i})$ in a neighborhood of any point in a cotangent bundle such that the symplectic 2-form acquires the canonical form $\omega = \dd x^{i} \wedge \dd p_{i}$ \cite{Vilasi}. However, on a Lie group it is more convenient and physically transparent to use the coordinates generated by the left-invariant (or right-invariant) vector fields, since they respect the division of the cotangent bundle to its horizontal $SO(3)$ and vertical $\mathfrak{so}(3)$ parts, i.e., to the configuration and the momentum spaces.} $P_{i} \equiv P_{g} \cdot T_{i,g}$ are Euclidean coordinates in the momentum space $\mathfrak{so}(3) \cong \mathbb{R}^{3}$. The Hamiltonian vector field $X_{F} \in \mathcal{T}(\mathcal{T}^{*}SO(3))$ corresponding to a function $F \in C^{1}(SO(3))$ is defined through the relation $i_{X_{F}}\omega \equiv \dd F$, where $i_{X}$ denotes the interior product, which ensures that $\mathcal{L}_{X_{F}}\omega = 0$, i.e., the symplectic structure is conserved under the flow generated by $X_{F}$. Then, the Poisson bracket $\{\cdot,\cdot\}_{PB}$ on the phase space $\mathcal{T}^{*}SO(3)$ is given by
\be
	\{F,G\}_{PB} \equiv \omega(X_{F},X_{G}) = \frac{\prt F}{\prt P_{i}} \frac{\prt G}{\prt g^{i}} - \frac{\prt F}{\prt g^{i}} \frac{\prt G}{\prt P_{i}} - 2\lo^{-1}\epsilon_{ij}^{\phantom{ij}k}P_{k} \frac{\prt F}{\prt P_{i}} \frac{\prt G}{\prt P_{j}} \pc
\ee
for $F,G \in C^{1}(\mathcal{T}^{*}SO(3))$, where we use the notation $\mathcal{L}_{(\lo^{-1}T_{i},0)}F =: \frac{\prt F}{\prt g^{i}}$ and $\mathcal{L}_{(0,\lo T^{i})}F =: \frac{\prt F}{\prt P_{i}}$ for the Lie derivation of a function $F \in C^{1}(\mathcal{T}^{*}SO(3))$ with respect to left-invariant basis vector fields in $\mathcal{T}(\mathcal{T}^{*}SO(3))$. Here we observe that the momentum space receives a non-trivial Poisson structure due to the non-commutativity of the left-invariant derivations under the Lie bracket.

\subsection{Parametrization of the phase space}
Although the Poisson brackets can be expressed in a global form in terms of Lie derivatives, and while the coordinate functions $P_{i} \equiv P_{g} \cdot T_{i,g}$ in the cotangent spaces are globally well-defined due to the parallelizability of $\mathcal{T}^{*}SO(3)$ with the left-invariant 1-forms $P_{g}$ labelled by $P_{i}$, the choice of coordinates on $SO(3)$ is not unique. In fact, since $SO(3)$ is compact, we cannot cover it with only one coordinate patch. For example, the coordinate functions $X^{i}(g)$, defined via $g \equiv \exp[iX^{i}(g) t_{i}/\lo]$, on the group are well-defined and differentiable only in a finite neighborhood of the unit element, namely, for $|X^{i}(g)| < \pi\lo/2$ (in the Euclidean $\mathbb{R}^{3}$ norm). For $|X^{i}(g)| = \pi\lo/2$ there is a two-to-one relation between a group element $g$ and $\pm X^{i}(g)$, since in this case $\exp[iX^{i}(g)t_{i}/\lo] = g = \exp[-iX^{i}(g)t_{i}/\lo]$. However, we can choose a coordinate system $X_{h}^{i}(g)$ in a neighborhood of any element $h \in SO(3)$, such that $g \equiv h\, \exp[iX_{h}^{i}(g)t_{i}/\lo]$. 

The Poisson brackets for the coordinate functions above, induced by the left-invariant vector fields, read
\be\label{eq:xppoisson}
	\{X_{h}^{i},X_{h}^{j}\}_{PB} = 0 \quad , \quad \{X_{h}^{i},P_{j}\}_{PB}\Big|_{h} = -\delta^{i}_{j} \quad , \quad \{P_{i},P_{j}\}_{PB} = -2\lo^{-1}\epsilon_{ij}^{\phantom{ij}k}P_{k}
\ee
for all $h \in SO(3)$. Note that the exact form of the Poisson bracket $\{X_{h}^{i},P_{j}\}_{PB}$ holds only at the origin of the coordinate system $h \in SO(3)$, but this is generally true for any choice of coordinates $Z_{h}(g)$ such that $\{Z_{h}^{i},P_{j}\}_{PB}\big|_{h} = -\delta^{i}_{j}$. However, the above choice of coordinates is not the most convenient one, since the inverse map from the group elements to the coordinate functions includes a logarithm, and makes it therefore difficult to make explicit calculations. Possibly a better choice are the coordinates 
\be
	Y_{h}^{i}(g) := -i\frac{\lo}{2}\tr(h^{-1}g\sigma^{i}) \equiv X_{h}^{i}(g) \frac{\sin(|X_{h}(g)|/\lo)}{|X_{h}(g)|/\lo} \pc
\ee
which are one-to-one for the range $|Y_{h}(g)| < \lo$. These coordinates will also later enter the definition of the non-commutative plane waves in Subsection \ref{sec:groupfourier}. For them the Poisson brackets read
\be\label{eq:yppoisson}
	\{Y_{h}^{i},Y_{h}^{j}\}_{PB} = 0 \quad , \quad \{Y_{h}^{i},P_{j}\}_{PB} = -\sqrt{1 - \lo^{-2}Y_{h}^{2}}\ \delta^{i}_{j} + \lo^{-1}Y_{h}^{k}\epsilon_{ki}^{\phantom{ki}j} \quad , \quad \{P_{i},P_{j}\}_{PB} = -2\lo^{-1}\epsilon_{ij}^{\phantom{ij}k}P_{k}
\ee
for the whole range of validity of $Y_{h}^{i}$, and we also have $\{Y_{h}^{i},P_{j}\}_{PB}\big|_{h} = -\delta^{i}_{j}$ as a special case. The deformation of the canonical commutation relations can be understood to arise from the deformed addition of the coordinates on $SO(3)$, given by
\bea
	\left( Y_{h}(g_{1}) \oplus_{h} Y_{h}(g_{2}) \right)^{i} & \equiv & Y_{h}^{i}(g_{1}h^{-1}g_{2}) \nonumber\\
	&=& \epsilon_{h}(g_{1},g_{2})\Big[ \sqrt{1 - Y_{h}^{2}(g_{2})/\lo^{2}}\ Y_{h}^{i}(g_{1}) + \sqrt{1 - Y_{h}^{2}(g_{1})/\lo^{2}}\ Y_{h}^{i}(g_{2}) \nonumber\\
	&& \qquad \qquad - \big(Y_{h}(g_{1}) \times Y_{h}(g_{2})\big)^{i}/\lo \Big]\ \mt{,} \label{eq:deformedadd}
\eea
where $\epsilon_{h}(g_{1},g_{2}) := \sgn(\tr(h^{-1}g_{1}h^{-1}g_{2}))$ in the fundamental $SU(2)$ representation and $\times$ denotes the Euclidean cross-product. This is because the momenta correspond to the generators of translations on the configuration space $SO(3)$. Furthermore, we observe that the Poisson algebra of the momentum space coordinate functions $P_{i}$ reflects directly the non-commutative structure of the $\mathfrak{so}(3)$ Lie algebra.

\subsection{Classical dynamics on $\mathcal{T}^{*}SO(3)$}
The dynamics of the system is given by the choice of a Hamiltonian function $H \in C^{1}(\mathcal{T}^{*}SO(3))$. The Hamiltonian vector field corresponding to $H$
\be
	\frac{\dd}{\dd t} \equiv {\{H, \cdot \}}_{PB} = \frac{\prt H}{\prt P_{i}} \frac{\prt}{\prt g^{i}} - \frac{\prt H}{\prt g^{i}} \frac{\prt}{\prt P_{i}} - 2\lo^{-1}\epsilon_{ij}^{\phantom{ij}k} P_{k} \frac{\prt H}{\prt P_{i}} \frac{\prt}{\prt P_{j}}
\ee
is the generator of time-evolution of dynamical variables in the phase space. This expression has an additional term with respect to the Euclidean case, arising from the non-commutative Lie algebra structure of derivations, which vanishes in the commutative limit $\lo \rightarrow \infty$. Thus the classical dynamics of the system coincides with the usual Euclidean one in this limit. In particular, the equations of motion for the canonical variables are
\bea\label{eq:Heoms}
	\frac{\dd X_{h}^{i}}{\dd t}\Big|_{h} & \equiv & {\{H, X_{h}^{i} \}}_{PB}\Big|_{h} = \frac{\prt H}{\prt P_{i}}\Big|_{h}\ \forall\ h \in SO(3) \pc\\
	\frac{\dd P_{i}}{\dd t} & \equiv & {\{H, P_{i} \}}_{PB} = - \frac{\prt H}{\prt g^{i}} + 2\lo^{-1}\epsilon_{ij}^{\phantom{ij}k} P_{k} \frac{\prt H}{\prt P_{j}} \pt
\eea

\subsection{The phase space of Loop Quantum Gravity}
Let us digress briefly to recall some elements of Loop Quantum Gravity \cite{Thiemann, carlo}, showing the similarity with the much simpler case we are dealing with here, as an additional motivation for our forthcoming analysis. In the connection formulation that defines the classical starting point of LQG, the configuration space of General Relativity is replaced, after a canonical transformation, by the space of $\mathfrak{so}(3)$-valued (or $\mathfrak{su}(2)$-valued, depending on the formulation) connections $A$ on some spatial 3d hypersurface $\Sigma$ in spacetime. The conjugate variable is a (densitized) triad field $e$, also valued in $\mathfrak{so}(3)$. One then changes variables in the classical phase space, going from connections to $SO(3)$-valued holonomies $h_l \equiv \mathcal{P}\, \exp[\int_{l}A]$ along paths $l$ in $\Sigma$, where $\mathcal{P}$ indicates path ordering. The canonically conjugate variable to the holonomies along all such possible paths are $\mathfrak{so}(3)$-valued fluxes $E_S \equiv \int_{S_{l}} *(e \wedge e)$ of the triad fields through surfaces, also embedded in $\Sigma$. More generally, one considers (spin network) graphs embedded in $\Sigma$ and associated dual surfaces \cite{Thiemann}.
Computing Poisson brackets of these classical variables, one finds commutativity among any two holonomies and non-commutativity among fluxes. While the general Poisson brackets are complicated, as they depend on the specific path and surface associated to the variables considered, in the simplest case of a link $l$ and a surface $S$ intersecting it at a single point, one finds the algebraic structure of $\mathcal{T}^{*}(SO(3)) \cong SO(3) \times \mathfrak{so}(3)$, the cotangent bundle over $SO(3)$  \cite{noncomm},
\[
\{h_l,h_l'\}=0 \quad , \quad \{E^i_S,E^j_{S'}\}=-2\delta_{S,S'}{\epsilon^{ij}}_k E^k_S \pc
\]
where $-2{\epsilon^{ij}}_k$ are the structure constants of the Lie algebra $\mathfrak{so}(3)$. Thus, the phase space $\mathcal{T}^{*}SO(3)$ we are considering in this paper is the phase space associated, in LQG, to a single link of  a spin network graph, with associated dual surface.

For a more general spin network $\Gamma$, the phase space $\mathcal{P}_{\Gamma}$ will become the direct product of phase spaces of the individual links $l \in \Gamma$ modulo relations arising from the requirement of $SO(3)$ gauge invariance at the vertices $v \in \Gamma$,
\be
	\mathcal{P}_{\Gamma} \equiv \mathcal{T}^{*}\left( SO(3)^L / SO(3)^V \right) \pt
\ee
where $L$ is the number of links in $\Gamma$ and $V$ is the number of vertices.
In terms of the fluxes $E_S$ the gauge invariance can be understood as implying the closure of the surfaces dual to the links attached to the vertex, or equivalently the absence of flux outgoing from the region surrounded by the same surfaces around the vertex.

Obviously, there is one crucial difference between our simple case of non-relativistic quantum mechanics on $SO(3)$ and LQG. In LQG, being the quantization of General Relativity, we must require diffeomorphism invariance to hold, and so due to the reparametrization invariance of the time coordinate, the Hamiltonian function $H$ becomes a constraint \cite{Thiemann}, and no time coordinate will play a distinguished role. Therefore, with regard to LQG, we restrict our arguments mainly on the level of phase space structure, and comment on the possible advantages and perspectives for LQG of the non-commutative structure we study, based on the intuition gained from the simpler cases.

\section{Quantum mechanics on $SO(3)$}\label{sec:qm}

\subsection{Group basis}\label{sec:groupbasis}
Building on the results reviewed in the previous section, it is now straightforward to give a canonical formulation of quantum mechanics on $SO(3)$. The natural choice for canonical variables in the phase space $SO(3) \times \mathfrak{so}(3)$ are the group elements $g \in SO(3)$ and the Lie algebra elements $P \in \mathfrak{so}(3) \cong \mathbb{R}^{3}$, which we then `upgrade' to canonical operators formally via the quantization map $\ \hat{}\ $: $g \mapsto \hat{g}$ and $P_{i} \mapsto \hat{P}_{i}$ for the components of $P \in \mathfrak{so}(3)$. We first consider the complete set of orthonormal basis states $\{ |g\rangle\ |\ g \in SO(3) \}$, labelled by group elements, which we will call the group basis, satisfying
\be
	\langle g | g' \rangle \equiv \frac{1}{\pi^{2}\lo^{3}}\delta(g^{-1}g') \quad , \quad \int_{SO(3)}\!\!\pi^{2}\lo^{3}\dd g\ |g\rangle \langle g | \equiv \hat{\1} \quad\mt{, and}\quad f(\hat{g})|g\rangle \equiv f(g)|g\rangle
\ee
for any function $f \in F(SO(3))$ over $SO(3)$, where $\hat{\1}$ denotes the unit operator, $\dd g$ the normalized Haar measure. In the above sense, $|g\rangle$ are the eigenstates of the operator $\hat{g}$. Note that here one must necessarily consider the group elements themselves as operators instead of some coordinate functions on $SO(3)$, since there doesn't exist any global differentiable coordinate system on $SO(3)$. On the other hand, since $g \in SO(3)$ are not real numbers but group elements, we can neither act directly with the operator $\hat{g}$ on states. The property $f(\hat{g})|g\rangle \equiv f(g)|g\rangle$, however, guarantees that for \emph{any coordinate system} on $SO(3)$, $|g\rangle$ are the eigenstates of the coordinates. As usual, we then define the Hilbert space of states $\mathcal{H}$ to consist of those states $|\psi\rangle$, whose decomposion in the $|g\rangle$ basis can be expressed in the form
\be
	|\psi\rangle = \int_{SO(3)}\!\!\pi^{2}\lo^{3}\dd g\ \psi(g)|g\rangle \pc
\ee
where $\psi \in L^{2}(SO(3),\pi^{2}\lo^{3}\dd g)$, and $\langle \psi | \psi \rangle = 1$. Notice that, with this choice of normalization, the measure coincides with the Lebesgue measure on $\mathbb{R}^3$ in the commutative limit $\lo\rightarrow\infty$. 

In the group basis $\{ |g\rangle\ |\ g \in SO(3) \}$ it is natural to choose the momentum operators to be of the form
\be
	\hat{P}_{i} \equiv -i\frac{\hbar}{l_{\circ}}\mathcal{L}_{T_{i}} \pc
\ee
where $\mathcal{L}_{T_{i}}$ is the Lie derivative on $SO(3)$ with respect to the left-invariant basis vector field $T_{i} \in \mathcal{T}SO(3)$, and $\hbar$ is the Planck constant, since these are the generators of translations on $SO(3)$. Also, they are self-adjoint in $L^{2}(SO(3),\pi^{2}\lo^{3}\dd g)$, and the quadratic Casimir operator is $\hat{P}^{2} := \sum_{i} \hat{P}_{i}^{2} \equiv -(\hbar/\lo)^{2} \Delta_{LB}$, where $\Delta_{LB}$ is the Laplace-Beltrami operator on $SO(3)$ \cite{Helgason}.

The commutators of the operators $\hat{P}_{i}$ and the coordinate operators $\hat{X}_{h}^{i} := X_{h}^{i}(\hat{g})$ corresponding to the coordinate functions $X_{h}^{i}: SO(3) \rightarrow \mathbb{R}^{3}$, induced by the left-invariant vector fields, $g \equiv h\, \exp[iX_{h}^{i}(g) t_{i}/l_{\circ}]$, where $t_{i} \equiv T_{i}(e) \in \mathcal{T}_{e}SO(3) \cong \mathfrak{so}(3)$ is a Lie algebra basis, and $|X_{h}(g)| < \pi\lo/2$, read
\be\label{eq:comrel}
	[\hat{X}_{h}^{i},\hat{X}_{h}^{j}] = 0 \quad , \quad {[\hat{X}_{h}^{i},\hat{P}_{j}]|}_{h} = i\hbar \delta_{j}^{i} \quad , \quad [\hat{P}_{i},\hat{P}_{j}] = 2i\frac{\hbar}{l_{\circ}}\epsilon_{ij}^{\phantom{ij}k}\hat{P}_{k} \pc
\ee
for all $h \in SO(3)$, due to the properties of Lie derivatives. As for the Poisson brackets, we find that, in general, we must have different canonical coordinate operators $\hat{X}_{h}^{i}$ for each $h \in SO(3)$ to satisfy the canonical commutation relations with the momentum operators. With the choices above, the commutators then correctly reflect the classical Poisson structure (\ref{eq:xppoisson}) of the canonical variables in the usual sense:
\be
	\frac{i}{\hbar}[\hat{O},\hat{O}'] \stackrel{\hbar \rightarrow 0}{\longrightarrow} \{O,O'\}_{PB} \pt
\ee

The canonical dynamics of a quantum system corresponding to a given classical system is then obtained by applying the quantization map to the Hamiltonian function, $H(P,g) \mapsto \hat{H} \equiv H(\hat{P},\hat{g})$. As is well known, this prescription is not unique due to the ambiguity in ordering operators. Namely, there are several inequivalent ways to order the canonical operators in $\hat{H}$, which all coincide in the limit $\hbar \rightarrow 0$. For the purposes of this paper, however, we will not have to fix any specific operator ordering. (Later in considering the first order phase space path integral we will restrict to consider Hamiltonians of the form $H(P,g) = H_{P}(P) + H_{g}(g)$ to avoid unnecessary complications arising from operator ordering ambiguities.) Nevertheless, the ordering should be such that the Hamiltonian operator $\hat{H}$ is self-adjoint, $\langle \psi' | H \psi \rangle = \langle H \psi' | \psi \rangle$. Then, the time-evolution generated by $\hat{H}$ is unitary, and we obtain a unitary time-evolution operator $\hat{U}(t'-t) := \exp[-\frac{i}{\hbar}(t'-t)\hat{H}]$.

\

In LQG, one proceeds to quantization in the same way as we have done above (apart from the normalization chosen for the measure of integration, which is taken to be simply the Haar measure on each link). Functions $\psi_\Gamma$ associated to graphs $\Gamma$ define the Hilbert space $\mathcal{H}_\Gamma\simeq L^2\left( SO(3)^L / SO(3)^V \right)$. By taking the direct sum over all such Hilbert spaces associated to all possible graphs embedded in $\Sigma$, one defines the kinematical Hilbert space of the theory. Similarly to what we have done above, one defines the algebra of kinematical observables of the theory by turning the canonical variables into operators.

\subsection{Spin basis and coherent states}\label{sec:spins}
In this Subsection we will briefly review the well-known decomposition of functions on $SO(3)$ in terms of spin representations, as well as in terms of coherent states on $SO(3)$. This will allow us to compare our results in later parts of the paper to the ones obtained by more conventional methods.

A decomposition of functions on $SO(3)$ in terms of spin representations follows from Peter-Weyl theorem \cite{VilenkinKlimyk}, which shows the orthogonality of the elements of representation matrices (also called Wigner D-functions) $D^{j}_{mn}(g)$, $j \in \mathbb{N}_{0}$, $m,n=0,\pm 1,\ldots,\pm j$:
\be
	\int_{SO(3)}\dd g\ \overline{D^{j'}_{m'n'}(g)} D^{j}_{mn}(g) = (2j+1)^{-1} \delta^{jj'} \delta_{mm'} \delta_{nn'} \pt
\ee
Accordingly, the states $|j;m,n\rangle$ defined via $\langle g | j;m,n \rangle := \sqrt{2j+1}\ D^{j}_{nm}(g)$ constitute a basis in the Hilbert space $\mathcal{H} \cong L^{2}(SO(3))$, the spin basis, for they satisfy
\be
	\langle j';m',n'|j;m,n \rangle = \pi^{2}\lo^{3}\delta^{j'j}\delta_{m'm}\delta_{n'n} \mand \frac{1}{\pi^{2}\lo^{3}}\sum_{j,m,n} |j;m,n\rangle \langle j;m,n| = \hat{\1} \pc \label{jmn}
\ee
where in the spin basis we introduce the measure $(\pi^{2}\lo^{3})^{-1} \sum_{j,m,n}$, $j \in \mathbb{N}_{0}$, $m,n=0,\pm 1, \ldots, \pm j$. The spin label $j$ of the state $|j;m,n\rangle$ refers to the eigenvalues of the quadratic Casimir operator $\hat{P}^{2} = (-i\hbar/\lo)^{2} \sum_{i} \mathcal{L}_{T_{i}} \mathcal{L}_{T_{i}}$,
\be
	\hat{P}^{2}|j;m,n\rangle = (2\hbar/\lo)^{2}j(j+1)|j;m,n\rangle \pc
\ee
which label the irreducible unitary representations of $SO(3)$, while the indices $m,n$ refer to eigenvalues of the Lie derivative operators $\hat{P}_{z}^{l,r} := -i\frac{\hbar}{\lo}\mathcal{L}_{T_{z}^{l,r}}$ with respect to left- and right-invariant vector fields $T^{l,r}_{z} \in \mathcal{T}SO(3)$, such that $T^{l}_{z}(e) = T^{r}_{z}(e) \equiv t_{z} \in \mathfrak{so}(3)$,
\be
	\hat{P}_{z}^{l}|j;m,n\rangle = 2\frac{\hbar}{\lo}m|j;m,n\rangle \mand \hat{P}_{z}^{r}|j;m,n\rangle = 2\frac{\hbar}{\lo}n|j;m,n\rangle \pc
\ee
respectively.\footnote{The factors of $2$ in these expressions follow from our choice of normalization of the basis vector fields, namely, we chose $T^{l}_{i}(e) = T^{r}_{i}(e) \equiv t_{i} = \sigma_{i}$ in the fundamental representation, where $\sigma_{i}$ are the Pauli matrices, for later convenience. With the choice $t_{i} = \sigma_{i}/2$ we would find expressions without extra factors of $2$.} These operators form a maximal commuting set of operators on the Hilbert space $\mathcal{H}$. The commutativity of left- and right-invariant derivatives follows from the associativity of the group product. Also note that the Casimir operators corresponding to left- and right-invariant Lie derivatives coincide. Then, any state $|\psi\rangle \in \mathcal{H}$ can be expanded in the spin basis as
\be
	| \psi \rangle = \frac{1}{\pi^{2}\lo^{3}}\sum_{j,m,n} |j;m,n\rangle \langle j;m,n | \psi \rangle \pc
\ee
where
\be
	\langle j;m,n | \psi \rangle = \int_{SO(3)}\pi^{2}\lo^{3}\dd g\ \langle j;m,n | g \rangle \langle g | \psi \rangle \equiv \int_{SO(3)}\pi^{2}\lo^{3}\dd g\ \sqrt{2j+1}\ \overline{D^{j}_{nm}(g)} \psi(g) \pt
\ee
We should emphasize that the states $|j;m,n\rangle$ are not the usual spherical harmonics states $|j,m\rangle$ familiar from considerations of quantum spin systems \cite{BrinkSatchler}, even though they are related through the definition of representation matrix elements $D^{j}_{mn}(g) \equiv \langle j,m | D^{j}(g) | j,n \rangle$. The spherical harmonics given by
\be
	\langle g | j,m\rangle := Y^{j}_{m}(g) \equiv \sqrt{(2j+1)/4\pi}\ \overline{D^{j}_{m0}(g)}
\ee
are invariant under the right multiplication $R_{h}^{*}f(g) \equiv f(gh)$ by elements of the form $e^{ist_{z}} \in U(1) \subset SU(2)$, and therefore are functions on $SU(2)/U(1) \cong \mathbb{S}^{2}$, the 2-sphere. In other words, the spherical harmonics $|j,m\rangle$ are elements of the vector space corresponding to the $(2j+1)$-dimensional representation, whereas we have states on the group itself. Thus, the spherical harmonics do not give a complete basis for $\mathcal{H} \cong L^{2}(SO(3))$, and we need to use the states $|j;m,n\rangle$ instead, which give a full representation of both left and right multiplication.

A set of coherent states\footnote{These states are directly related to Perelomov coherent states $|j,\vec{n}\rangle_{P}$ \cite{Perelomov} on a 2-sphere $\mathbb{S}^{2}$ via $\langle g | j;\vec{m},\vec{n}\rangle \equiv {}_{P}\langle j,\vec{m}|D^{j}(g)|j,\vec{n}\rangle_{P}$, the use of which has proven very useful in recent developments in Spin Foam models \cite{eterasimone,eterasimone2,ConradyFreidel}.} $|j;\vec{m},\vec{n}\rangle$, $j \in \mathbb{N}_{0}$, $\vec{m},\vec{n} \in \mathbb{S}^{2}$ (the unit 2-sphere in $\mathbb{R}^{3}$), can be defined on $SO(3)$ in terms of the representation matrices as
\be
	\langle g |j;\vec{m},\vec{n}\rangle := \sqrt{\frac{2j+1}{\pi^{2}\lo^{3}}}\ D^{j}_{jj}(g_{\vec{n}}^{-1}gg_{\vec{m}}) \pc \label{coh}
\ee
where $g_{\vec{m}} \in SO(3)$ is the unique element, which rotates the unit vector in $z$-direction in $\mathbb{R}^{3}$ to $\vec{m}$ with the axis of rotation in the $x$-$y$-plane (similarly for $\vec{n}$). These states satisfy $\hat{P}^{2}|j;\vec{m},\vec{n}\rangle = 4(\hbar/\lo)^{2}j(j+1)|j;\vec{m},\vec{n}\rangle$,
\be
	\vec{m}\cdot\hat{P}^{l}|j;\vec{m},\vec{n}\rangle = 2\frac{\hbar}{\lo}j|j;\vec{m},\vec{n}\rangle \mand \vec{n}\cdot\hat{P}^{r}|j;\vec{m},\vec{n}\rangle = 2\frac{\hbar}{\lo}j|j;\vec{m},\vec{n}\rangle \pc
\ee
but also
\be
	\langle j;\vec{m},\vec{n}|\hat{P}^{l}_{i}|j;\vec{m},\vec{n}\rangle = 2\frac{\hbar}{\lo}j\vec{m}_{i} \mand \langle j;\vec{m},\vec{n}|\hat{P}^{r}_{i}|j;\vec{m},\vec{n}\rangle = 2\frac{\hbar}{\lo}j\vec{n}_{i} \pt
\ee
The coherent states $|j;\vec{n},\vec{n}\rangle =: |j;\vec{n}\rangle$ are particularly interesting, because they turn out to coincide in the semi-classical limit $\hbar \rightarrow 0$, $j \rightarrow \infty$, $\hbar j = const.$, with the states $|P\rangle$ defined below in terms of the non-commutative momentum variables $P \in \mathbb{R}_{\star}^{3}$ in the sense that, at this limit, we can identify the variables $P = 2(\hbar/\lo)j \vec{n}$. This suggests that the variable $2(\hbar/\lo)j\vec{n} \in \mathfrak{so}(3)$ labelling the states $| j; \vec{n}\rangle$ can be understood in some sense as `quantum momentum variables' for a particle on $SO(3)$, between the continuous variables $P$ and the discrete quantum numbers $(j,m,n)$ in character, while the functions $\langle g | j; \vec{n}\rangle = \sqrt{(2j+1)/\pi^{2}\lo^{3}}\ D^{j}_{jj}(g_{\vec{n}}^{-1}gg_{\vec{m}}) $ may be understood as a sort of plane waves giving rise to an alternative Fourier transform (given a suitable non-commutative product for the variable $\vec{n} \in \mathbb{S}^{2}$) to the one we will use in the following \cite{carlos}.

\

The above decompositions correspond, in Loop Quantum Gravity, to the expansion of cylindrical functions of the connection in spin network basis \cite{Thiemann, carlo}, which are labelled by quantum numbers of geometric operators, functions of the triad operator (areas, volumes, etc.), in the same sense in which the states (\ref{jmn}) are eigenstates of the quantum operators corresponding to the momentum variables $P$.

\subsection{Group Fourier transform}\label{sec:groupfourier}
Next we will review the formulation of group Fourier transform on $SO(3)$ considered first in \cite{FreidelLivine,FreidelMajid} (and extended to $SU(2)$ in \cite{JoungMouradNoui}) in connection with 3d quantum gravity models, and has also been utilized in formulating a non-commutative flux representation of Loop Quantum Gravity \cite{NCflux} and a metric representation of Group Field Theory \cite{BaratinOriti,GFTdiffeos}. On the most foundational level, the transform is based on the theory of quantum groups \cite{majid}, but we will restrict our review only on the most pragmatic aspects. For more details, see \cite{FreidelMajid,JoungMouradNoui}.

\

The group Fourier transform on $SO(3)$ is an isometry \cite{NCflux} in $L^{2}$-norm from $L^{2}$-functions on $SO(3)$ to $L^{2}$-functions on a non-commutative space $\mathbb{R}_{\star}^{3}$ equipped with a $\star$-product. To describe the transform and the non-commutative space $\mathbb{R}_{\star}^{3}$, we first define the non-commutative plane waves,
\be
	e_{g}(P) := \exp[\tr(gP)l_{\circ}/2\hbar] \equiv \exp[iY^{i}(g)P_{i}/\hbar]
\ee
on $SO(3)$, where the trace $\tr(gP)$ is taken in the fundamental 2-dimensional representation of $SO(3)$, which is obtained from $g' \in SU(2)$ via the 2-to-1 map $g' \mapsto \sgn(\tr(g'))g' \equiv g \in SO(3)$. $e_{g}(P)$ will act in the construction as the kernel of the group Fourier transform. The coordinates $Y^{i}(g)$ in the exponential are given by
\be
	Y^{i}(g) \equiv \frac{\lo}{2}\tr(g\sigma^{i}) \equiv X^{i}(g) \frac{\sin(|X(g)|/\lo)}{|X(g)|/\lo} \pc
\ee
where $g \equiv \exp[iX^{i}(g)t_{i}/\lo]$. The non-commutative $\star$-product is then defined through the relation\footnote{Strictly speaking, this definition of the $\star$-product of non-commutative plane waves makes only sense under integration: We can derive for any involutive group element 
\be
	g^{2}=e\ \Rightarrow\ e_{g^{2}}(X) = 1\ \Rightarrow\ e_{g}(X) = e_{g^{-1}}(X) \pc
\ee
where to get the last equality we $\star$-multiplied both sides by $e_{g^{-1}}(X)$. This last equality, however, is not true in general for the involutive elements of $SO(3)$ with the above chosen form for the plane waves. Nevertheless, by excluding the involutive elements, which constitute a set of measure zero on $SO(3)$, from the integration range of the group Fourier transform, we can make rigorous sense of this definition in the context of the transform.}

\be\label{eq:starprod}
	e_{g}(P) \star e_{h}(P) \equiv e_{gh}(P) \qquad \forall\ g,h \in SO(3),\ P \in \mathfrak{so}(3) \pt
\ee
Obviously, the $\star$-product so defined is associative but non-commutative due to the corresponding properties of the group multiplication. It also induces a corresponding deformed non-commutative addition \cite{FreidelLivine2} for the coordinates $Y^{i}(g)$ on $SO(3)$ such that $\exp[iY_{1} \cdot P/\hbar] \star \exp[iY_{2} \cdot P/\hbar] \equiv \exp[i(Y_{1} \oplus Y_{2}) \cdot P/\hbar]$, where $\oplus \equiv \oplus_{e}$ from (\ref{eq:deformedadd}).

Now, we may define the group Fourier transform $\tilde{f} \in L^{2}(\mathbb{R}_{\star}^{3})$ of a function $f \in L^{2}(SO(3))$ as
\be
	\tilde{f}(P) := \int_{SO(3)}\!\! \pi^{2}\lo^{3}\dd g\ \overline{e_{g}(P)} f(g) \pc
\ee
where $\overline{e_{g}(P)}$ denotes the complex conjugate of $e_{g}(P)$. The inverse transform is then given by
\be
	f(g) \equiv \int_{\mathbb{R}_{\star}^{3}} \frac{\dd^{3} P}{(2\pi\hbar)^{3}}\ e_{g}(P) \star \tilde{f}(P) \pc
\ee
where $\dd^{3} P$ is the usual Lebesgue measure on $\mathbb{R}^{3}$. In particular, we have the important identities
\be\label{eq:deltas}
	\int_{\mathbb{R}_{\star}^{3}} \frac{\dd^{3} P}{(2\pi\hbar)^{3}}\ e_{g}(P) = \frac{1}{\pi^{2}\lo^{3}}\delta(g) \mand \int_{SO(3)}\!\! \pi^{2}\lo^{3}\dd g\ \overline{e_{g}(P)} = (2\pi\hbar)^{3}\delta_{\star}(P) \pc
\ee
where $\delta_{\star}$ is the delta function with respect to the $\star$-product in the sense
\be
	\int_{\mathbb{R}_{\star}^{3}} \dd^{3} P\ \delta_{\star}(P) \star \tilde{f}(P) = \tilde{f}(0) = \int_{\mathbb{R}_{\star}^{3}} \dd^{3} P\ \tilde{f}(P) \star \delta_{\star}(P) \pt
\ee
We should note that $\delta_{\star}(P)$ is a regular function on $\mathbb{R}^{3}$ for $\hbar\lo^{-1} > 0$, and in particular $\delta_{\star}(0) < \infty$. It is also easy to show that the group Fourier transform is an isometry \cite{NCflux} with respect to the $L^{2}$-norms:
\be
	\int_{SO(3)}\!\! \pi^{2}\lo^{3}\dd g\ \overline{f(g)}f'(g) = \int_{\mathbb{R}_{\star}^{3}} \frac{\dd^{3} P}{(2\pi\hbar)^{3}}\ \overline{\tilde{f}(P)} \star \tilde{f}'(P) \pt
\ee
Another useful property of the $\star$-product, of which we will take advantage later on, is that it can be expressed as a differential operator under integration \cite{FreidelLivine2,Livine} via
\be\label{eq:starint}
	\int \dd^{3}P\ f(P) \star g(P) = \int \dd^{3}P\ f(P) \sqrt{1 + (\hbar/\lo)^{2} \Delta_{P}}\ g(P) \pc
\ee
where $\Delta_{P} := \sum_{j} \frac{\prt}{\prt P_{j}}\frac{\prt}{\prt P^{j}}$ is the Laplacian in $\mathbb{R}^{3}$ with respect to the momentum variable $P$.

In fact, the map $e: SO(3) \rightarrow \{ e_{g}\ |\ g \in SO(3) \}$, $g \stackrel{e}{\mapsto} e_{g}$ is an isomorphism: $gh \stackrel{e}{\mapsto} e_{gh} \equiv e_{g} \star e_{h}$, the inverse $e^{-1}: \{ e_{g}\ |\ g \in SO(3) \} \rightarrow SO(3)$ to the fundamental representation being
\be\label{eq:ginv}
	e_{g} \stackrel{e^{-1}}{\mapsto} \left. \left( \1_{2} \sqrt{1 + \left(\frac{\hbar}{l_{\circ}}\right)^{2}\Delta_{P}} + \frac{\hbar}{l_{\circ}}\sigma_{j}\frac{\prt}{\prt P_{j}} \right) e_{g}\right|_{P=0} = g \pc
\ee
where $\1_{2}$ is the $2 \times 2$ identity matrix, and $\sigma_{j}$ are the Pauli matrices. Then, the non-commutative plane waves $e_{g}$ give an `isomorphic' representation of $SO(3)$ on functions $f \in L^{2}(\mathbb{R}_{\star}^{3})$ via $\star$-multiplication as
\be
	(e_{g} \star f)(P) = \int_{SO(3)}\!\! \pi^{2}\lo^{3}\dd h\ \overline{e_{h}(P)} f(gh) \in L^{2}(\mathbb{R}_{\star}^{3}) \pc
\ee
which is dual to left translations $L_{g}^{*} f(h) = f(gh) \in L^{2}(SO(3))$. Right translations will instead be dual to $(f \star e_{g})(P)$.

Moreover, a definition of $\star$-polynomials of the coordinates $P_{i}$ in the non-commutative space (which do not belong to $L^{2}(\mathbb{R}_{\star}^{3})$) and their $\star$-products with the non-commutative plane waves can be obtained through the relation
\bea
	(-i\frac{\hbar}{l_{\circ}})^{n} \big( \mathcal{L}_{T_{i_{1}}} \cdots \mathcal{L}_{T_{i_{n}}} e_{g}(P) \big) & = & e_{g}(P) \star (-i\frac{\hbar}{l_{\circ}}) \mathcal{L}_{T_{i_{1}}}e_{e}(P) \star \cdots \star (-i\frac{\hbar}{l_{\circ}})\mathcal{L}_{T_{i_{n}}}e_{e}(P) \nonumber\\
	& \equiv & e_{g}(P) \star P_{i_{1}} \star \cdots \star P_{i_{n}} \label{eq:starpoly}\pc
\eea
from which we obtain
\be
	P_{i_{1}} \star \cdots \star P_{i_{n}} = e_{g}(P) \star P_{i_{1}} \star \cdots \star P_{i_{n}}\Big|_{g=e} = (-i\frac{\hbar}{l_{\circ}})^{n} \big( \mathcal{L}_{T_{i_{1}}} \cdots \mathcal{L}_{T_{i_{n}}} e_{g}(P) \big)\Big|_{g=e} \pt
\ee
With this definition, we find that the lowest order term in $\hbar\lo^{-1}$ is, for all powers, the point-wise product,
\be
	P_{i_{1}} \star \cdots \star P_{i_{n}} = P_{i_{1}} \cdots P_{i_{n}} + \mathcal{O}(\hbar/\lo) \pt
\ee
Thus, the (non-commutative) $\star$-product coincides with the (commutative) point-wise product both in the classical limit $\hbar \rightarrow 0$ as well as in the Euclidean (commutative) limit $\lo \rightarrow \infty$. We also find, by an explicit calculation,
\be\label{eq:pstarsquared}
	P_{i} \star P_{j} = P_{i} P_{j} + i\frac{\hbar}{\lo}\epsilon_{ij}^{\phantom{ij}k}P_{k} \pc
\ee
so the coordinates $P_{i}$ satisfy Lie algebra type of commutation relations, which follow directly from the properties of the Lie derivative. Moreover, we have for the non-commutative plane waves an expression in terms of the $\star$-product as
\bea
	e_{g}(P) & \equiv & e^{X^{i}(g) \mathcal{L}_{T_{i}}/\lo} e_{e}(P) \nonumber\\
	& = & \sum_{n=0}^{\infty} \frac{1}{n!\lo^{n}} X^{i_{1}} \cdots X^{i_{n}} (\mathcal{L}_{T_{i_{1}}} \cdots \mathcal{L}_{T_{i_{n}}} e_{e}(P)) \nonumber\\
	& = & \sum_{n=0}^{\infty} \frac{i^{n}}{n!\hbar^{n}} X^{i_{1}} \cdots X^{i_{n}} P_{i_{1}} \star \cdots \star P_{i_{n}} \nonumber\\
	& = & e_{\star}^{i X(g) \cdot P / \hbar} \label{eq:starwave} \pc
\eea
where $g \equiv \exp[iX^{i}(g)t_{i}/\lo]$, $t_{i} \equiv T_{i,e}$, $|X(g)| \leq \pi\lo/2$, and we introduced the notation
\be
	e_{\star}^{\lambda f(P)} := \sum_{n=0}^{\infty}\frac{\lambda^{n}}{n!} \underbrace{f(P) \star \cdots \star f(P)}_{n\ \mathrm{times}} \pt
\ee

\subsection{The non-commutative momentum basis}\label{sec:momentumbasis}
Through the group Fourier transform reviewed in the previous section, we may now formulate the dual momentum space representation of quantum mechanics on $SO(3)$ in terms of the non-commutative momentum space $\mathbb{R}_{\star}^{3}$. (See, e.g., \cite{Madore,FSK} for earlier treatments of quantum mechanics on Lie algebraic non-commutative spaces.) Let us define a set of states $\{ |P\rangle\ |\ P \in \mathbb{R}_{\star}^{3} \}$ via their inner product with the group basis
\be
	\langle g | P \rangle \equiv e_{g}(P) \pc
\ee
which satisfy, due to the properties (\ref{eq:deltas}) of the group Fourier transform, the following formal identities
\be
	\langle P | P' \rangle = (2\pi\hbar)^{3}\delta_{\star}(P - P') \mand \int_{\mathbb{R}_{\star}^{3}} \frac{\dd^{3} P}{(2\pi\hbar)^{3}}\ |P \rangle \star \langle P | = \hat{\1} \pt
\ee
Accordingly, they form a basis with respect to the $\star$-product structure in the non-commutative momentum space. 

The momentum operators $\hat{P}_{i}$ do not commute among themselves but obey the Lie algebra commutation relations in (\ref{eq:comrel}), and therefore it is impossible to diagonalize them simultaneously. However, the states $|P\rangle$ are eigenstates of the momentum operators $\hat{P}_{i}$ in the $\star$-product sense $\hat{P}_{i} |P\rangle = |P\rangle \star P_{i}$: Due to (\ref{eq:starpoly}), by definition,
\be
	\langle g | \hat{P}_{i} | P \rangle \equiv -i \frac{\hbar}{l_{\circ}} \mathcal{L}_{T_{i}} e_{g}(P) \equiv e_{g}(P) \star P_{i} \equiv \langle g | P \rangle \star P_{i} \pt
\ee
where $\lo^{-1}T_{i}$ are the left-invariant basis vector fields, and the ordering of $P_{i}$ to the right of $\langle g | P \rangle$ follows from the left-invariance of $\lo^{-1}T_{i}$. For the operators $\hat{P}^{r}_{i} \equiv -i\frac{\hbar}{\lo}\mathcal{L}_{T^{r}_{i}}$ corresponding to right-invariant basis vector fields $T^{r}_{i}$ we have $\langle g | \hat{P}^{r}_{i} | P \rangle = P_{i} \star \langle g | P \rangle $. Since the $\star$-product is non-commutative, and thus non-local in the sense that it is not possible to resolve points, the states $|P\rangle$ are, from the point of view of commutative $\mathbb{R}^{3}$, coherent states peaked on the momentum value $P$: $\langle P | \hat{P}_{i} | P \rangle =  P_{i} \langle P | P \rangle = P_{i} \delta_{\star}(0)$.\footnote{Manipulating formal expressions in the Dirac bra-ket notation requires some extra care now that we have two different products in use. In particular, when evaluating expressions of the form $\langle P | \hat{O} | P \rangle$, where $\hat{O}$ is some operator, one should always first evaluate $\langle P | \hat{O} | P' \rangle$ and only afterwards set $P=P'$, i.e., we define $\langle P | \hat{O} | P \rangle := \langle P | \hat{O} | P' \rangle |_{P=P'}$.}

By the properties of the group Fourier transform, any state $|\psi\rangle \in \mathcal{H}$ has then a representation in terms of the momentum basis as
\be
	|\psi\rangle = \int_{\mathbb{R}_{\star}^{3}} \frac{\dd^{3} P}{(2\pi\hbar)^{3}}\ |P\rangle \star  \tilde{\psi}(P) \pc
\ee
where $\tilde{\psi}(P) \equiv \langle P | \psi \rangle \in L^{2}(\mathbb{R}_{\star}^{3})$ is the group Fourier transform of the wave function $\psi(g) \equiv \langle g | \psi \rangle \in L^{2}(SO(3))$ in the group basis. The group element operator $\hat{g}$ in the fundamental representation can be written in the momentum basis, due to (\ref{eq:ginv}), as
\be
	\hat{g} = \1_{2} \sqrt{1 + \frac{\hbar^{2}}{l_{\circ}^{2}}\Delta_{P}} + \frac{\hbar}{l_{\circ}}\sigma_{j}\frac{\prt}{\prt P_{j}} \pc
\ee
where $\1_{2}$ is the $2 \times 2$ unit matrix and $\sigma_{j}$ are the Pauli matrices.

Since the transformations between different bases are isometries, we may always freely move from one basis to another. In particular, the transformation between the spin basis and the non-commutative momentum basis is mediated by the transformation kernel \cite{Livine,HanThiemann}
\bea
	\langle j;m,n | P \rangle &=& \int_{SO(3)}\!\!\pi^{2}\lo^{3}\dd g\ \sqrt{2j+1}\ \overline{D^{j}_{nm}(g)} e_{g}(P) \nonumber\\
	&=&   \pi^{2}\lo^{3} \sqrt{2j+1} \frac{2J_{2j+1}(\lo|P|/\hbar)}{i^{2j}\lo|P|/\hbar} D^{j}_{mn}(e^{i\frac{\pi\lo}{2\hbar}\frac{P}{|P|}}) \pc
\eea
In the semi-classical limit $\hbar \rightarrow 0$, $j \rightarrow \infty$, $\hbar j = const.$ the function $2J_{2j+1}(\lo|P|/\hbar)/i^{2j}\hbar^{-1}\lo|P|$ approximates a delta function $\delta(|P| - 2\frac{\hbar}{\lo}j)$ \cite{Livine}.

Moreover, as also noted in \cite{NCflux}, for the coherent states $|j;\vec{n}\rangle$ it can be shown that in the semi-classical limit $\hbar \rightarrow 0$, $j \rightarrow \infty$, $\hbar j = const.$, their group Fourier transform
\be
	\langle P | j;\vec{n} \rangle = \sqrt{\pi^{2}\lo^{3}(2j+1)} \int_{SO(3)} \dd g\ \overline{e_{g}(P)} D^{j}_{jj}(g_{\vec{n}}^{-1}gg_{\vec{n}})
\ee
peaks sharply at $P = 2(\hbar/\lo)j\vec{n}$, and therefore we find an identification of the non-commutative dual variables and the coherent state variables in this limit, in which, we remark again, the momentum variables $P$ become commutative. This can be further understood by first noting that for the states $|P\rangle$ we have $\hat{P}^{2}|P\rangle = P^{2} \star |P\rangle$,
\be
	\hat{P}^{l}_{i}|P\rangle = |P\rangle \star P_{i} \mand \hat{P}^{r}_{i}|P\rangle = P_{i} \star |P\rangle \pc
\ee
while for the coherent states $|j;\vec{n}\rangle$ we have $\hat{P}^{2}|j;\vec{n}\rangle = \left(2\hbar/\lo\right)^{2}j(j+1)|j;\vec{n}\rangle$ and
\be
	\langle j;\vec{n}| \hat{P}^{l}_{i}|j;\vec{n}\rangle = 2\frac{\hbar}{\lo}j = \langle j;\vec{n}| \hat{P}^{r}_{i}|j;\vec{n}\rangle \pt
\ee
Accordingly, in the limit $\hbar/\lo \rightarrow 0$, $j \rightarrow \infty$, $(\hbar/\lo)j = const.$, where the $\star$-product coincides with the point-wise product and the coherent states become infinitely sharply peaked, both $|P\rangle$ and $|j;\vec{n}\rangle$ are simultaneous eigenstates (in the commutative sense) of the maximal set of commuting operators $\hat{P}^{2}$, $\hat{P}^{l,r}_{i}$ with the above identification of the eigenvalues.

\

At this point, before moving onto the path integral formulation, we would like to further highlight the connection between the non-trivial classical Poisson structure (\ref{eq:xppoisson}) of the momentum variables, the canonical commutation relations (\ref{eq:comrel}) of the canonical quantum operators, and the non-commutative structure of the momentum space induced by the $\star$-product (\ref{eq:starprod}). These all arise from a common source, namely, the non-commutative Lie algebra structure of derivations on $SO(3)$, which again follows from, or rather is equal to, the curvature of the group manifold. There are certain choices we have made in deriving these structures. The first choice is the specific coordinate system for the phase space, which is reflected in the canonical Poisson structure (\ref{eq:xppoisson}) of the classical theory. We chose to use the coordinates induced by the left-invariant vector fields on $SO(3)$, which allow for unique global coordinates for the covector spaces, and accordingly for a single Euclidean momentum space. In this respect, it is the most natural choice of coordinates even though, according to the Darboux theorem, one could always choose a \emph{local} coordinate system such that one recovers the usual Poisson brackets, but these coordinates mix the coordinates of configuration and momentum spaces, and cannot be globally extended. The second choice is the choice of quantum mechanical momentum operators in the group basis such that they correctly reflect the classical Poisson structure. We chose to use the Lie derivatives with respect to the left-invariant vector fields on $SO(3)$. These are a natural generalization of the directional coordinate derivatives $\frac{\prt}{\prt x^{i}}$ in the Euclidean case as the generators of translations in $SO(3)$, and indeed coincide with them at the Euclidean limit $\lo \rightarrow 0$. Also, this choice is consistent with the fact that we chose to use the coordinates generated by the left-invariant vector fields for the configuration space. Furthermore, these operators are self-adjoint with respect to the measure we use on $SO(3)$, and the quadratic operator $\hat{P}^{2} = \sum_{i}\hat{P}_{i}\hat{P}_{i}$ is a multiple of the Laplace-Beltrami operator on $SO(3)$. The third choice has to do with the non-commutative plane waves, and thus the $\star$-product structure, which is not completely unique. In particular, it seems that one can choose any infinitely differentiable real-valued coordinates $X^{j}(g)$ on $SO(3)$ for the plane waves $e_{g}(P) \equiv \exp[iP_{j}X^{j}(g)/\hbar]$, which satisfy $X^{j}(g^{-1}) = -X^{j}(g)$ and $\mathcal{L}_{T_{i}}X^{j}(e) = \delta_{i}^{j}$. In general, this will affect the explicit form of the $\star$-product but, however, not the Lie algebraic commutation relations of the momentum variables, since these derive directly from the commutation relations of the Lie derivatives via the relation (\ref{eq:starpoly}).

\

 In LQG, an analogous non-commutative basis was introduced in \cite{NCflux}, and interpreted as providing a flux representation for the theory, taking into account the non-commutativity of fluxes even in the classical theory \cite{noncomm}.  We close by noting that the same quantization map we use here, and related to the Duflo map \cite{FreidelMajid}, has been used, and justified further, in the context of Chern-Simons path integrals in \cite{HannoThomas}.

\subsection{Path integral in terms of non-commutative variables}\label{sec:pathintegral}
Next we will give the first order path integral formulation of quantum mechanics on $SO(3)$ using the non-commutative momentum space defined above, and in particular the momentum basis $\{ |P\rangle\ |\ P \in \mathfrak{so}(3)\}$. The derivation follows similar lines to the commutative case, but some extra subtleties arise due to the non-commutative structure.

The quantum mechanical evolution operator is given by
\be
	\hat{U}(t'-t) \equiv e^{-\frac{i}{\hbar}(t'-t)\hat{H}} \pc
\ee
where $\hat{H}$ is the Hamiltonian operator, as usual. Accordingly, we have for the propagation amplitude from the group element $g$ at time $t$ to $g'$ at time $t'$
\be
	\langle g',t' | g,t \rangle \equiv \langle g' | \hat{U}(t'-t) | g \rangle \pt
\ee
Now, we introduce the time-slicing via the decomposition
\be
	\hat{U}(t'-t) \equiv \prod_{k=0}^{N-1} \hat{U}(t_{k+1}-t_{k}) \pc
\ee
where $t_{k+1} > t_{k}\ \forall k$ and $t_{0} = t, t_{N} = t'$. We set $t_{k+1} - t_{k} \equiv \epsilon\ \forall\ k=0,\ldots,N-1$, so we have $t'-t \equiv N\epsilon$. By inserting the resolution of identity $\hat{\1} = \int_{SO(3)}\pi^{2}\lo^{3}\dd g\ |g\rangle \langle g|$ $N-1$ times in between the evolution operators we obtain
\be
	\langle g',t' | g,t \rangle = \lim_{N \rightarrow \infty} \left[ \prod_{k=1}^{N-1} \int_{SO(3)}\!\!\pi^{2}\lo^{3}\dd g_{k} \right] \left[ \prod_{k=0}^{N-1} \langle g_{k+1} | \hat{U}(\epsilon) | g_{k} \rangle \right] \pt
\ee
Furthermore, for each of the factors we use the resolution of identity $\hat{\1} = \int_{\mathbb{R}_{\star}^{3}} \frac{\dd^{3} P}{(2\pi\hbar)^{3}}\ |P\rangle \star \langle P|$ to express them as
\be
	\langle g_{k+1} | \hat{U}(\epsilon) | g_{k} \rangle = \int_{\mathbb{R}_{\star}^{3}} \frac{\dd^{3} P_{k}}{(2\pi\hbar)^{3}} \langle g_{k+1} | P_{k}\rangle \star \langle P_{k} | \hat{U}(\epsilon) | g_{k} \rangle \pt
\ee
At this point, we have to restrict to systems with Hamiltonian operators of the form $\hat{H} = H_{P}(\hat{P}) + H_{g}(\hat{g})$ to avoid additional operator ordering issues, on top of those stemming from the non-commutativity of momentum variables and encoded by the $\star$-product structure defined above.\footnote{To handle more general Hamiltonians with mixed terms in $g$ and $P$ variables, one should introduce an additional $\star$-product, encoding in the definition of the path integral the operator ordering between group and momentum operators \cite{ChaichianDemichev}. This would then lead to more complicated forms for the discrete and continuum phase space path integrals. However, for the arguments we wish to present in this paper, the generalization is not important, as we focus on how the $\star$-product between momentum variables encodes their non-commutativity in the same path integral representation of the dynamics, so we restrict to the simpler case.} Then we get $\langle P | \hat{H} | g \rangle = H_{\star}(P,g) \star \langle P | g \rangle$, where the function $H_{\star}(P,g)$ is now obtained from the Hamiltonian operator $\hat{H}$ by replacing the momentum operators $\hat{P}_{i}$ in $\hat{H}$ by the non-commutative momentum variables $P_{i}$ and the operator product of the momentum operators is replaced by the $\star$-product, whereas the group element operators $\hat{g}$ are replaced by the group elements $g$ and the operator product between them by the group multiplication. We may take the linear approximation in $\epsilon$ as
\be
	\langle P | e^{-\frac{i}{\hbar} \epsilon \hat{H}} | g \rangle \approx \big(1 - \frac{i}{\hbar} \epsilon H_{\star}(P,g) \big) \star \langle P | g \rangle \approx e_{\star}^{-\frac{i}{\hbar} \epsilon H_{\star}(P,g)} \star \langle P | g \rangle \pc
\ee
since the linear order in $\epsilon$ for the time-slice propagators is sufficient in order to obtain the correct finite time propagator satisfying the Schr\"odinger equation \cite{DeWitt}. Accordingly, we obtain
\bea
	\langle g_{k+1} | \hat{U}(\epsilon) | g_{k} \rangle &\approx& \int_{\mathbb{R}_{\star}^{3}} \frac{\dd^{3} P_{k}}{(2\pi\hbar)^{3}} e_{g_{k+1}}(P_{k}) \star e_{\star}^{-\frac{i}{\hbar} \epsilon H_{\star}(P_{k},g_{k})} \star e_{g_{k}^{-1}}(P_{k}) \nonumber\\
	&=& \int_{\mathbb{R}_{\star}^{3}} \frac{\dd^{3} P_{k}}{(2\pi\hbar)^{3}} e_{g_{k+1}g_{k}^{-1}}(P_{k}) \star e_{\star}^{-\frac{i}{\hbar} \epsilon H_{\star}(g_{k}^{-1}P_{k}g_{k},g_{k})} \pc
\eea
where we used the properties $\overline{e_{g}(P)} = e_{g^{-1}}(P)$,
\be
	e_{g_{k+1}}(P) \star e_{g_{k}^{-1}}(P) = e_{g_{k+1}g_{k}^{-1}}(P_{k}) \mand e_{g}(P) \star f(P) = f(g^{-1}Pg) \star e_{g}(P)
\ee
of the $\star$-product. Furthermore, we have $e_{g_{k+1}g_{k}^{-1}}(g_{k} P_{k} g_{k}^{-1}) = e_{g_{k}^{-1}g_{k+1}}(P_{k})$, so by making the change of variables $P_{k} \mapsto g_{k}P_{k}g_{k}^{-1}$ the full propagator reads in the first order form
\bea
	\langle g',t' | g,t \rangle &=& \lim_{N \rightarrow \infty} \left[ \prod_{k=1}^{N-1} \int_{SO(3)}\pi^{2}\lo^{3}\dd g_{k} \right] \left[ \prod_{k=0}^{N-1} \int_{\mathbb{R}_{\star}^{3}} \frac{\dd^{3} P_{k}}{(2\pi\hbar)^{3}} \right] \nonumber\\
	& & \qquad \times \left[ \prod_{k=0}^{N-1} e_{g_{k}^{-1}g_{k+1}}(P_{k}) \star e_{\star}^{-\frac{i}{\hbar} \epsilon H_{\star}(P_{k},g_{k})} \right] \pt \label{eq:gdiscpi}
\eea
We observe that each of the factors in the product over $P_{k}$'s is exactly the group Fourier transform of the function $\exp_{\star}[-\frac{i}{\hbar} \epsilon H_{\star}(P_{k},g_{k})]$ from the momentum variable $P_{k}$ to the group variable $g_{k}^{-1}g_{k+1}$. It is easy to verify that the time-slice propagator
\be
	\langle g_{k+1}| \hat{U}(\epsilon) | g_{k} \rangle = \int_{\mathbb{R}_{\star}^{3}} \frac{\dd^{3} P_{k}}{(2\pi\hbar)^{3}} e_{g_{k}^{-1}g_{k+1}}(P_{k}) \star e_{\star}^{-\frac{i}{\hbar} \epsilon H_{\star}(P_{k},g_{k})}
\ee
satisfies the Schr\"odinger equation exactly. However, we would like to express this as an integral over a single exponential. Using the expression (\ref{eq:starint}) for the $\star$-product under integration, and taking again the linear approximation in $\epsilon$, we obtain
\be
	\langle g_{k+1}| \hat{U}(\epsilon) | g_{k} \rangle = \int_{\mathbb{R}_{\star}^{3}} \frac{\dd^{3} P_{k}}{(2\pi\hbar)^{3}}\ \exp\left\{ \frac{i}{\hbar} \epsilon \left[ \frac{Y(g_{k}^{-1}g_{k+1})}{\epsilon} \cdot P_{k} - H_{q}(P_{k},g_{k}) \right] \right\} \pc
\ee
where
\be\label{eq:qHam}
	H_{q}(P,g) := \sqrt{1 + \left(\hbar/\lo\right)^{2} \Delta_{P}}\ H_{\star}(P,g)
\ee
is an effective Hamiltonian containing additional terms, which arise from the non-trivial phase space structure and ensure that the time-slice propagator satisfies the Schr\"odinger equation up to first order in $\epsilon$. Accordingly, we may write
\bea
	\langle g',t' | g,t \rangle &=& \lim_{N \rightarrow \infty} \left[ \prod_{k=1}^{N-1} \int_{SO(3)}\pi^{2}\lo^{3}\dd g_{k} \right] \left[ \prod_{k=0}^{N-1} \int_{\mathbb{R}_{\star}^{3}} \frac{\dd^{3} P_{k}}{(2\pi\hbar)^{3}} \right] \nonumber\\
	& & \qquad \times \exp\left\{ \frac{i}{\hbar} \sum_{k=0}^{N-1} \epsilon \left[ \frac{Y(g_{k}^{-1}g_{k+1})}{\epsilon} \cdot P_{k} - H_{q}(P_{k},g_{k}) \right] \right\} \pc \label{eq:discpi}
\eea
This is clearly analogous to the usual first order form of path integral in the usual Euclidean case. The second order path integral can, in principle, be obtained from this first order form by integrating out the momentum or group variables, but since these integrations can only be performed explicitly for certain special cases (quadratic Hamiltonians, in particular), for greater generality we will stay at the first order level. (See Subsection \ref{sec:freeparticle} for the free particle on $SO(3)$ case.)

We may write (\ref{eq:discpi}) in the continuum limit as
\be\label{eq:gpi}
	\langle g',t' | g,t \rangle = \int_{\substack{g(t)=g\\g(t')=g'}} \mathcal{D}g\ \mathcal{D}P\ \exp \left\{ \frac{i}{\hbar} \int_{t}^{t'}\!\! \dd s\ \Big[ \frac{\lo}{2} \tr\big(\dot{g}(s) P(s)\big) - H_{q}(P(s),g(s)) \Big] \right\} \pc
\ee
where $\dot{g}(t) := -i g^{-1}(t) \frac{\dd g}{\dd t}(t) \in \mathfrak{so}(3)$, since we have
\be
	\lim_{\epsilon \rightarrow 0} \frac{Y(g_{k}^{-1}g_{k+1})/\lo}{\epsilon} = -i \left. \frac{\dd}{\dd \epsilon} \right|_{\epsilon = 0} g^{-1}(t_{k}) g(t_{k} + \epsilon) \equiv \dot{g}(t_{k}) \pt
\ee
In the expression (\ref{eq:gpi}) of the propagator, the action
\be
	\mathcal{S}_{b}[g,P] =\int_{t}^{t'}\!\! \dd s\ \Big[ \frac{\lo}{2} \tr\big(\dot{g}(s) P(s)\big) - H_{q}(P(s),g(s)) \Big]
\ee
appearing in the exponent is the classical action, i.e., the time integral over the classical Lagrangean function obtained through Legendre transformation, except for the effective Hamiltonian $H_{q}$. The Hamiltonian $H_{q}$ can be interpreted as introducing quantum corrections to the action, as it contains, in addition to the classical Hamiltonian function $H$ in the zeroth order, higher order terms in $\hbar$. 
The presence of such quantum corrections to the classical action in the path integral formulation of the dynamics is necessary in order for the propagator to satisfy the Schr\"odinger equation, and a generic feature of path integrals on curved manifolds \cite{DeWitt,ChaichianDemichev}.
Also, note that at the Euclidean (no curvature in configuration space, commutative in momentum space) limit $\lo \rightarrow \infty$ we have $H_{q} \rightarrow H$ and, as should be expected, at this limit the path integral (\ref{eq:gpi}) coincides with the path integral for a point particle in Euclidean space.

The expression (\ref{eq:gpi}) represents our first main result, confirming the usefulness and interpretation of the non-commutative momentum basis, in this simple context.

We should also mention that the homotopy group of $SO(3)$ is $\mathbb{Z}_{2}$ and, therefore, in the path integral the paths are divided into two separate homotopy classes, which may receive different phase factors \cite{Dowker2}. The phases must, however, constitute a unitary representation of $\mathbb{Z}_{2}$, and there are altogether two choices, the multiplicative groups $\{1\}$ and $\{1,-1\}$, which give two different propagators. It can be shown \cite{Schulman} that the first one corresponds to the integer spin representations $j \in \mathbb{N}_{0}$ while the latter corresponds to the half-integer spin representations $j \in \mathbb{N}_{0} + \frac{1}{2}$ of $SO(3)$.

\

The propagator $\langle P',t' | P,t \rangle$ in the non-commutative momentum basis is obtained by applying the group Fourier transform to both sides of the propagator $\langle g',t' | g,t \rangle$ in the group basis. This results in adding a boundary term into the action:
\bea
	\langle P',t' | P,t \rangle &=& \int_{SO(3)} \pi^{2}\lo^{3}\dd g' \int_{SO(3)} \pi^{2}\lo^{3}\dd g\ \langle P' | g' \rangle \langle g',t' | g,t \rangle \langle g | P \rangle \nonumber\\
	&=& \int_{SO(3)} \pi^{2}\lo^{3}\dd g' \int_{SO(3)} \pi^{2}\lo^{3}\dd g\ e^{-\tr(g'P')\lo/2\hbar + \tr(gP)\lo/2\hbar}  \nonumber\\
	&& \times \int_{\substack{g(t)=g\\g(t')=g'}} \mathcal{D}g\ \mathcal{D}P\ \exp \left\{ \frac{i}{\hbar} \int_{t}^{t'}\!\! \dd s\ \Big[ \lo \tr\big(\dot{g}(s) P(s)\big) - H_{q}(P(s),g(s)) \Big] \right\} \nonumber\\
	&=& \int_{\substack{P(t)=P\\P(t')=P'}} \mathcal{D}g\ \mathcal{D}P\ \exp \Bigg\{ \frac{i}{\hbar} \int_{t}^{t'}\!\! \dd s\ \Big[ \frac{\lo}{2} \tr\big(\dot{g}(s) P(s)\big) - H_{q}(P(s),g(s)) \Big] \nonumber\\
	&& \qquad \qquad \qquad \qquad \quad - \frac{\lo}{2\hbar} \Big[ \tr\big( g(t')P(t') - \tr\big( g(t)P(t)\big) \Big] \Bigg\} \nonumber\\
	&\equiv& \int_{\substack{P(t)=P\\P(t')=P'}} \mathcal{D}g\ \mathcal{D}P\ e^{\frac{i}{\hbar}\mathcal{S}[g,P]}\pc \label{eq:pi}
\eea
where the action $\mathcal{S}[g,P]$ consists of bulk and boundary terms $\mathcal{S}[g,P] \equiv \mathcal{S}_{b}[g,P] + \mathcal{S}_{\prt b}[g,P]$, respectively,
\bea
	\mathcal{S}_{b}[g,P] &=& \int_{t}^{t'}\!\! \dd s\ \Big[ \frac{\lo}{2} \tr\big(\dot{g}(s) P(s)\big) - H_{q}(P(s),g(s)) \Big] \pc \nonumber\\
	\mathcal{S}_{\prt b}[g,P] &=& i\frac{\lo}{2} \Big[ \tr\big( g(t')P(t') - \tr\big( g(t)P(t)\big) \Big] \pt \label{eq:Paction}
\eea
Here, the appearance of a boundary term in the action is analogous to what happens in path integral quantization of BF theory, in which the $B$ field represents the analogue of our momentum variable $P$, and in general for gravity theories when one chooses  metric boundary conditions, i.e., fixes (some components of) the intrinsic metric on the boundary. This boundary term is similarly crucial for obtaining the correct semi-classical limit in our case, as we will observe in Subsection \ref{sec:semiclassicalanalysis}.

Let us summarize the results of this section. We have shown that one can derive a first order path integral for a quantum mechanical system with configuration space $SO(3)$ in terms of the non-commutative momentum space variables defined in Subsection \ref{sec:momentumbasis}. For the propagator in the group basis we obtained the continuum limit expression (\ref{eq:gpi}),
\be
	\langle g',t' | g,t \rangle = \int_{\substack{g(t)=g\\g(t')=g'}} \mathcal{D}g\ \mathcal{D}P\ \exp \left\{ \frac{i}{\hbar} \int_{t}^{t'}\!\! \dd s\ \Big[ \frac{\lo}{2} \tr\big(\dot{g}(s) P(s)\big) - H_{q}(P(s),g(s)) \Big] \right\} \pc
\ee
where the measure reads
\be
	\mathcal{D}g\ \mathcal{D}P\ \equiv \lim_{N \rightarrow \infty} \left[ \prod_{k=1}^{N-1} \pi^{2}\lo^{3}\dd g_{k} \right] \left[ \prod_{k=0}^{N-1} \frac{\dd^{3} P_{k}}{(2\pi\hbar)^{3}} \right] \pt
\ee
The $\star$-product structure gives naturally rise to quantum corrections into the action
\be
	\mathcal{S}_{b}[g,P] = \int_{t}^{t'}\!\! \dd s\ \Big[ \frac{\lo}{2} \tr\big(\dot{g}(s) P(s)\big) - H_{q}(P(s),g(s)) \Big]
\ee
via the form of the quantum corrected Hamiltonian $H_{q}(P,g) \equiv \sqrt{1 + (\hbar/\lo)^{2} \Delta_{P}}\ H_{\star}(P,g)$. Crucially, these corrections ensure that the propagator obtained via path integral satisfies the Schr\"odinger equation. In the momentum basis we found that the action receives an additional boundary term
\be
	\mathcal{S}_{\prt b}[g,P] = i\frac{\lo}{2} \Big[\tr\big( g(t')P(t') \big) - \tr\big( g(t)P(t) \big) \Big] \pc
\ee
and thus the path integral acquires the form (\ref{eq:pi}),
\bea
	\langle P',t' | P,t \rangle &=& \int_{\substack{P(t)=P\\P(t')=P'}} \mathcal{D}g\ \mathcal{D}P\ \exp \Bigg\{ \frac{i}{\hbar} \int_{t}^{t'}\!\! \dd s\ \Big[ \frac{\lo}{2} \tr\big(\dot{g}(s) P(s)\big) - H_{q}(P(s),g(s)) \Big] \nonumber\\
	&& \qquad \qquad \qquad \qquad \quad - \frac{\lo}{2\hbar} \Big[\tr\big( g(t')P(t') \big) - \tr\big( g(t)P(t) \big) \Big] \Bigg\} \pt
\eea
The second order path integral, either in terms of the group or the momentum variables, can be obtained from the first order formalism by integrating out the momentum or the group variables, respectively.


\

The first order path integral we have derived above has a counterpart in the quantum gravity case, in the context of Spin Foam models and Loop Quantum Gravity, where it corresponds to the simplicial gravity path integral in terms of flux (triad) and connection variables \cite{BaratinOriti,GFTdiffeos}. If we were to expand (\ref{eq:pi}) in representation basis (\ref{jmn}), or in the associated coherent states (\ref{coh}), we would obtain the particle analogue of the \textit{Spin Foam} formulation of the dynamics of Loop Quantum Gravity \cite{SF}, i.e., a definition of the transition amplitude in terms of quantum numbers of geometric operators. To be concrete, in the case of 3d gravity, in Euclidean signature and no cosmological constant (thus with the gauge group $SO(3)$), the Spin Foam formulation of the dynamics for a given 2-complex (history of spin networks) made of vertices $v$, links $l$ and faces $f$ gives
\bea Z(\Gamma)=\left(\prod_{f}\,\sum_{j_{f}}\right)\,\prod_{f}(2j_{f} +1)\,\prod_{v}\, \left\{ \begin{array}{ccc}
j_1 &j_2 &j_3
\\ j_4 &j_5 &j_6
\end{array}\right\} \pc \label{sf}
\eea
while the non-commutative metric formulation yields the expression
\be
	\label{bf} Z(\Gamma) =  \int \prod_l \dd h_l \prod_f \dd x_f  \, e^{i \sum_f \tr( x_f H_f(h_l))}
\ee
with Lie algebra variables $x_f$ playing the role of discretized triad, group elements $h_l$ for each link $l$ of the 2-complex interpreted as a discrete connection, with discrete curvature $H_f = \prod_{l\in\partial f} h_l$. (\ref{bf}) is the usual expression for the simplicial path integral of 3d gravity in first order form, i.e., with the continuum action $ S(e,\omega)\,=\,\int_{\mathcal{M}} \Tr \left( e\wedge F(\omega)\right) $  
with triad 1-form $e^i(x)$ and a connection 1-form $\omega^{j}(x)$, both valued in $\mathfrak{so}(3)$, and the associated curvature 2-form $F(\omega)$. Looking at the derivation of this simplicial path integral in \cite{BaratinOriti}, one may notice that the amplitude corresponds to a simple composition of non-commutative plane waves, so that the BF dynamics resembles the trivial dynamics $\hat{H}=0$ in the particle case. See \cite{SF,BaratinOriti} for more details. We are going to give another simple explicit example of this duality between path integral and state sum formulations of dynamics in the following.

\subsection{Semi-classical analysis}\label{sec:semiclassicalanalysis}
We are now interested in the semi-classical approximation of the transition amplitudes we have derived in general form. The way we tackle this issue is to first show that the path integral expression we have derived has the correct semi-classical approximation, which confirms further the analysis we have performed and the role of the non-commutative momentum variables, and then use this result for obtaining a semi-classical approximation for the same transition amplitude written in representation space.

\

So, let us first of all study the variations to the action (\ref{eq:Paction}).
We choose an arbitrary path $(\bar{g}(s),\bar{P}(s))$ in the phase space, and introduce a perturbation of the path as $(\bar{g}(s)e^{i\eta Z(s)},\bar{P}(s)+\xi Q(s))$, where $Z(s),Q(s) \in \mathfrak{so}(3)$ for $s \in [t,t']$, and we assume that the momentum variation vanishes at the boundary: $Q(t) = Q(t') = 0$. We find for the first order variation of the tangent vector $\dot{g} := -ig^{-1}\frac{\dd g}{\dd s}$ the form
\be
	\delta\dot{g}(s) = \eta \left(\frac{\dd Z}{\dd s}(s) + i[\dot{g},Z](s) \right) + \mathcal{O}(\eta^{2}) \in \mathfrak{so}(3) \pt
\ee
Now, even though we seem to be dealing with a classical action in calculating the variations, we cannot forget its quantum origin. In fact, what we are really dealing with are the underlying quantum amplitudes in which the action appears. Thus, we must take into account the non-commutative $\star$-product structure in calculating the variations by defining the variation of the action through the variation of the amplitude, symbolically as
\bea
	&& \exp\Big\{ \frac{i}{\hbar}\Big( \eta \frac{\delta\mathcal{S}}{\delta g}[\bar{g},\bar{P}] \delta g + \xi \frac{\delta\mathcal{S}}{\delta P}[\bar{g},\bar{P}] \delta P \Big) + \mathcal{O}(\eta^{2},\xi^{2},\eta\xi) \Big\} \nonumber\\
	&:=& \exp\Big\{-\frac{i}{\hbar}S[\bar{g},\bar{P}] \oplus \frac{i}{\hbar}S[\bar{g}e^{i\eta Z},\bar{P}+\xi Q]\Big\} \nonumber\\
	&=& \exp\Big\{-\frac{i}{\hbar}S[\bar{g},\bar{P}]\Big\} \star \exp\Big\{\frac{i}{\hbar}S[\bar{g}e^{i\eta Z},\bar{P}+\xi Q]\Big\} \pc
\eea
where the $\star$-product applies for momentum variables in the same time-slice. $\oplus$ is the induced deformed addition from (\ref{eq:deformedadd}). Here, the ordering of the factors follows from the left-invariance of vector fields.

Due to the linear approximation in $\epsilon$ for the single time-slice actions in the bulk, we can however neglect the deformation of the addition in the calculation of the variation of the bulk part of the action. Substituting the variations into the bulk action, we find for the first order variation in $\eta$ and $\xi$
\bea
	\delta\mathcal{S}_{b}[\bar{g},\bar{P}] &=& \int_{t}^{t'}\!\! \dd s\ \Bigg\{ \eta \frac{\lo}{2}  \tr\Big[\Big(\frac{\dd Z}{\dd s}(s) + i[\dot{g},Z](s) \Big) \bar{P}(s)\Big] + \xi \frac{\lo}{2} \tr\big(\dot{\bar{g}}(s) Q(s)\big) \nonumber\\
	& & \qquad \qquad - \eta Z^{i}(s) \mathcal{L}_{T_{i}} H_{q}(\bar{P}(s),\bar{g}(s)) - \xi Q_{i}(s) \frac{\prt H_{q}}{\prt P_{i}}(\bar{P}(s),\bar{g}(s)) \Bigg\} \nonumber\\
	&=& \lo \eta Z^{i}(s) \bar{P}_{i}(s) \Big|_{s=t}^{s=t'} \nonumber\\
	& & + \int_{t}^{t'}\!\! \dd s\ \Bigg\{ \eta Z^{i}(s) \Big[ -\lo \frac{\dd \bar{P}_{i}}{\dd s}(s) + 2\lo \epsilon_{ij}^{\phantom{ij}k} \dot{g}^{j}(s) \bar{P}_{k} - \mathcal{L}_{T_{i}} H_{q}(\bar{P}(s),\bar{g}(s)) \Big] \nonumber\\
	& & \qquad \qquad + \xi Q_{i}(s) \Big[ \lo \dot{\bar{g}}^{i}(s) - \frac{\prt H_{q}}{\prt P_{i}}(\bar{P}(s),\bar{g}(s)) \Big] \Bigg\} \pt \label{eq:1stvar}
\eea
Let us, at first, neglect the first term in (\ref{eq:1stvar}) associated with the boundary. By requiring the variation given by the integral to vanish for arbitrary perturbations $Z^{i}(s),Q^{i}(s)$, we obtain the equations
\bea
	\lo \dot{\bar{g}}^{i}(s) &=& \frac{\prt H_{q}}{\prt P_{i}}(\bar{P}(s),\bar{g}(s)) \nonumber\\
	 \frac{\dd \bar{P}_{i}}{\dd s}(s) &=& 2\epsilon_{ij}^{\phantom{ij}k} \dot{g}^{j}(s) \bar{P}_{k} - \lo^{-1} \mathcal{L}_{T_{i}} H_{q}(\bar{P}(s),\bar{g}(s)) \pt
\eea
Substituting the first equation into the second, using the notation $\lo^{-1} \mathcal{L}_{T_{i}} =: \frac{\prt}{\prt g^{i}}$ from Section \ref{sec:class}, and noting that $\lo \dot{g}^{i} \equiv \frac{\dd X_{g}^{i}}{\dd t}\big|_{g}$, we arrive at the equations
\bea
	\frac{\dd \bar{X}_{\bar{g}}^{i}}{\dd t}\Big|_{\bar{g}} &=& \frac{\prt H_{q}}{\prt P_{i}}\Big|_{\bar{g}} \nonumber\\
	\frac{\dd \bar{P}_{i}}{\dd t} &=& - \frac{\prt H_{q}}{\prt g^{i}} +  2\lo^{-1}\epsilon_{ij}^{\phantom{ij}k} \frac{\prt H_{q}}{\prt P_{j}} \bar{P}_{k} \label{eq:1var} \pt
\eea
In the semi-classical limit $\hbar \rightarrow 0$ the dominating contribution to the path integral arises then from the paths satisfying the equations of motion (\ref{eq:1var}). Given that at this limit $H_{q} \rightarrow H$, the equations coincide with the classical equations of motion (\ref{eq:Heoms}) we obtained from the canonical analysis in Section \ref{sec:class}.

We still need to show that the boundary term in the first order variation of the action (\ref{eq:1stvar}) is cancelled by the variation of the boundary term. Now, for the boundaries no approximation is available, such as the one in $\epsilon$ for the bulk, and therefore the deformation structure must be taken into account. On the other hand, we observe that it is exactly the deformation of addition of coordinates on $SO(3)$, which enables us to cancel the boundary term and arrive at the right classical equations of motion: With the deformed addition $\oplus$ of coordinates on $SO(3)$, we obtain for the first order variation in $\eta$ of the boundary action
\bea
	\delta \mathcal{S}_{\prt b}[\bar{g},\bar{P}] &=& \left. -i\frac{\lo}{2} \tr\big( \bar{g}(s)\bar{P}(s) \big) \right|_{s=t}^{s=t'} \oplus \left. i\frac{\lo}{2} \tr\big( \bar{g}(s)e^{i\eta Z(s)}\bar{P}(s) \big) \right|_{s=t}^{s=t'} \nonumber\\
	&=& \left. i\frac{\lo}{2} \tr\big( e^{i\eta Z(s)}\bar{P}(s) \big) \right|_{s=t}^{s=t'} \nonumber\\
	&\approx& \left. -\frac{\lo}{2} \tr\big( \eta Z(s)\bar{P}(s) \big) \right|_{s=t}^{s=t'} \nonumber\\
	&=& -\lo \eta Z(s) \cdot \bar{P}(s) \Big|_{s=t}^{s=t'} \pc
\eea
which exactly cancels the boundary term arising from the bulk action (\ref{eq:1stvar}). This further confirms the correctness of the non-commutative Fourier transform in encoding the relation between $g$ and $P$ variables, needed to produce the boundary term in the action.

\

Accordingly, we obtain the correct semi-classical behavior from the path integral (\ref{eq:pi}), but only by taking into account the non-commutative structure of the phase space. In particular, this means that in the semi-classical limit $\hbar \rightarrow 0$ we may approximate the full path integral by a sum over the amplitudes of solutions to the classical equations of motion,
\be
	\langle P',t' | P,t \rangle \approx \sum_{(g_{cl},P_{cl})} e^{\frac{i}{\hbar}\mathcal{S}[g_{cl},P_{cl}]} \pc
\ee
such that $P_{cl}(t) = P$, $P_{cl}(t') = P'$.

\

Finally, the expression of the path integral in terms of the non-commutative momentum variables may be particularly useful, since it provides a simpler way to compute the semi-classical expansion of the transition amplitudes, even when they are written in the spin basis $|j;m,n\rangle$, or with the coherent states $|j;\vec{n}\rangle$. This is indeed the type of calculation that is more commonly used in Spin Foam models \cite{ConradyFreidel, asymp}.  Let us show how the above results can be used to this end in our simpler context.

\

In the semi-classical limit $j \rightarrow \infty$, $\hbar \rightarrow 0$, while $\hbar j = const.$, the coherent states and the non-commutative momentum states coincide, $P = 2(\hbar/\lo)j\vec{n}$ \cite{Livine,NCflux}. Therefore, we may easily calculate the semi-classical limit of the propagator $\langle j',\vec{n}',t'| j,\vec{n},t \rangle$ by transforming to the problem into the non-commutative momentum basis,
\be
	\langle j',\vec{n}',t'| j,\vec{n},t \rangle = \int_{\mathbb{R}_{\star}^{3}}\!\! \frac{\dd^{3} P'}{(2\pi\hbar)^{3}}\!\! \int_{\mathbb{R}_{\star}^{3}}\!\! \frac{\dd^{3} P}{(2\pi\hbar)^{3}}\ \langle j',\vec{n}'|P'\rangle \langle P',t'| P,t \rangle \langle P | j,\vec{n}\rangle \pc
\ee
taking the limit $j,j' \rightarrow \infty$, $\hbar \rightarrow 0$, while $\hbar j,\hbar j' = const.$, of this expression, and evaluating the path integral at the solutions to the classical equations of motion (\ref{eq:Heoms}), thus obtaining
\be
	\langle j',\vec{n}',t'| j,\vec{n},t \rangle \approx \sum_{(g_{cl},P_{cl})} e^{\frac{i}{\hbar}\mathcal{S}[g_{cl},P_{cl}]}
\ee
for phase space paths $(g_{cl},P_{cl})$ such that $P_{cl}(t) = 2(\hbar/\lo)j\vec{n}$, $P_{cl}(t') = 2(\hbar/\lo)j'\vec{n}'$, i.e., simply taking the path integral result and substituting the coherent state parameters associated (in this approximation) to the corresponding momentum variables. In particular, if the action appearing in the path integral, when evaluated on saddle points, depends on the norm of $P$ only (as in the free particle case, for example), then one obtains an expression depending only on the spin label $j$.

\

Let us again use the results obtained above to get some insight into the quantum gravity context.
The above semi-classical analysis sheds some light on the \emph{a priori} surprising asymptotic expressions for Spin Foam models for gravity \cite{SF,asymp,ConradyFreidel}. These take the form of a simplicial path integral weighted by the Regge action for discrete gravity, even though the appearance of the spin foam amplitudes themselves is far from resembling a discrete gravity path integral. By reflecting on the above result, and keeping in mind the dual expression of the same spin foam amplitudes as simplicial path integrals in non-commutative metric variables \cite{BaratinOriti}, one realizes that this is not surprising at all: the spin foam amplitudes acquire the same functional dependence on the non-commutative metric variables in the limit in which the variables appearing in the spin foam expression $j,\vec{n}$ coincide with the non-commutative metric variables themselves.\footnote{This had been anticipated in \cite{NCflux}.}
As an explicit example, consider the famous asymptotic expression of the $6j$-symbol that is the basic building block of the amplitudes (\ref{sf}) in the semi-classical limit:
\be
\left\{ \begin{array}{ccc}
j_1 &j_2 &j_3
\\ j_4 &j_5 &j_6
\end{array}\right\} \approx \frac{ e^{i\sum_{f} \left( (2j_{f}\; +\; 1)\; \Theta_{f}/2 + \frac{\pi}{4}\right)} + e^{-i\sum_{f} \left((2j_{f}\; +\; 1)\; \Theta_{f}/2 + \frac{\pi}{4}\right)} }{\sqrt{3\pi V}} \pc
\ee
where the function $\sum_{f}( 2j_{f} + 1)\; \Theta_{f}(j)/2$ is nothing but the Regge calculus action for a simplicial complex made of a single tetrahedron with six edge lengths $j_f$ and six dihedral angles $\Theta_f(j)$, functions of the same edge lengths, and volume $V(j)$. This is simply the saddle point evaluation with respect to the discrete connection $h_l$ of the path integral (\ref{bf}), for the same simplicial complex, dual to a single vertex of the 2-complex $\Gamma$, in which therefore all the discrete triad variables $x_f$ are on the boundary (and thus not subject to variations), which in the semi-classical limit have length $j_f$.

\subsection{Free particle on $SO(3)$}\label{sec:freeparticle}
Finally, in order to show explicitly the compatibility of our analysis and results with those obtained by more conventional methods (in particular, harmonic analysis on the group manifold), we first compare our path integral expression for the finite time propagator with the standard expression, in the special case of a free particle on $SO(3)$, i.e., with Hamiltonian $\hat{H} = \hat{P}^{2}/2m$, a multiple of the Casimir operator.
Let us stress that this is indeed a very special case, as we will notice by the fact that the transition amplitude written in coherent state basis basically coincide in form, apart from the discreteness of $j$ labels as opposed to the continuous norm of the $P$ variables, with the first order path integral. As we remarked above, this is not at all the case in general (as it is not in Spin Foam models). Nevertheless, the free particle is an important test case for the formalism we have developed in this paper, since it is well-known from previous literature \cite{DeWitt,Schulman,Dowker,ChaichianDemichev}.

\

In the canonical picture, the evolution operator for a free particle reads $\hat{U}(t'-t) = \exp[-\frac{i}{\hbar}(t'-t)\hat{P}^{2}/2m]$, and using the spin basis $|j;m,n\rangle$ we obtain a simple expression for the finite time propagator $\langle g',t'| g,t \rangle := \langle g'| \hat{U}(t'-t) | g \rangle$ 
\bea
	\langle g',t'| g,t \rangle &=& \frac{1}{\pi^{2}\lo^{3}}\sum_{j,m,n} \langle g'|j;m,n\rangle \langle j;m,n | e^{-\frac{i}{\hbar}(t'-t)\frac{1}{2m}\hat{P}^{2}} | g \rangle \nonumber\\
	&=& \frac{1}{\pi^{2}\lo^{3}} \sum_{j} e^{-\frac{i}{\hbar}(t'-t)\frac{1}{2m}\left(\frac{2\hbar}{\lo}\right)^{2} j(j+1)} (2j+1) \sum_{m} D^{j}_{mm}(g'g^{-1}) \nonumber\\
	&=& \frac{1}{\pi^{2}\lo^{3}} \sum_{j} e^{-\frac{i}{\hbar}(t'-t)\frac{2\hbar^{2}}{m\lo^{2}}j(j+1)} (2j+1) \chi^{j}(g'g^{-1}) \label{eq:spinprop}
\eea
in terms of the character function $\chi^{j}(g) := \tr D^{j}(g)$. Since the Hamiltonian is just the Casimir operator, the propagator is invariant under left and right translations of the elements $g,g'$, and depends only on the norm $|X(g'g^{-1})|$ of the coordinates on $SO(3)$ (defined via $g \equiv e^{iX^{i}(g)t_{i}}$) as \cite{Varshalovich}
\be
	\chi^{j}(g'g^{-1}) = \frac{\sin\left((2j+1)|X(g'g^{-1})|/\lo\right)}{\sin(|X(g'g^{-1})|/\lo)} \pc
\ee
and we may write
\be\label{eq:angleprop}
	\langle g',t'| g,t \rangle = \frac{1}{\pi^{2}\lo^{3}} \sum_{j} e^{-\frac{i}{\hbar}(t'-t)\frac{2\hbar^{2}}{m\lo^{2}}j(j+1)} (2j+1) \frac{\sin\left((2j+1)|X(g'g^{-1})|/\lo\right)}{\sin(|X(g'g^{-1})|/\lo)} \pt
\ee
This is, of course, the well-known evolution kernel \cite{Camporesi}.
In the limit $(t'-t) \rightarrow 0$ the propagator correctly satisfies $\langle g'| \hat{U}(t'-t) | g \rangle \rightarrow \frac{1}{\pi^{2}\lo^{3}} \delta(g'g^{-1})$. 

\

Now, for comparison, we perform the same calculation in the momentum basis $|P\rangle$, and we obtain
\bea
	\langle g',t'| g,t \rangle &=& \int_{\mathbb{R}_{\star}}\frac{\dd^{3}P}{(2\pi\hbar)^{3}} \langle g' | P \rangle \star \langle P | e^{-\frac{i}{\hbar}(t'-t)\frac{1}{2m}\hat{P}^{2}} | g \rangle \nonumber\\
	&=& \int_{\mathbb{R}_{\star}}\frac{\dd^{3}P}{(2\pi\hbar)^{3}} e_{\star}^{-\frac{i}{\hbar}(t'-t)\frac{1}{2m} \sum_{i} P_{i} \star P_{i}} \star e_{g'g^{-1}}(P) \pt
\eea
Here only the plane wave depends on the direction $\frac{P}{|P|} =: \vec{n} \in \mathbb{S}^{2}$ of $P$, since $\sum_{i} P_{i}\star P_{i} \equiv P^{2}$, so we may write
\bea
	\langle g',t'| g,t \rangle &=& \frac{1}{(2\pi\hbar)^{3}} \int_{\mathbb{R}_{+}} \dd |P|\ e_{\star}^{-\frac{i}{\hbar}(t'-t)\frac{1}{2m}P^{2}} \star \int_{\mathbb{S}^{2}} P^{2} \dd^{2}\vec{n}\  e_{g'g^{-1}}(|P|\vec{n}) \nonumber\\
	&=& \frac{1}{(2\pi\hbar)^{3}} \int_{\mathbb{R}_{+}} \dd |P|\ e_{\star}^{-\frac{i}{\hbar}(t'-t)\frac{1}{2m}P^{2}} \star 4\pi |P| \frac{\sin(|P||Y(g'g^{-1})|/\hbar)}{|Y(g'g^{-1})|/\hbar} \pt
\eea
Note that $P^{2}$ $\star$-commutes with everything, so there is no ambiguity in placing it in the above expression, where it arises from the integration measure. Given that $|Y(g'g^{-1})| = \lo \sin(|X(g'g^{-1})|/\lo)$, we can further write
\be
	\langle g',t'| g,t \rangle = \frac{1}{\pi^{2}\lo^{3}} \int_{\mathbb{R}_{+}} \frac{\lo}{2\hbar} \dd |P|\ e_{\star}^{-\frac{i}{\hbar}(t'-t)\frac{1}{2m}P^{2}} \star \frac{\lo}{\hbar}|P| \frac{\sin\left(\frac{\lo}{\hbar}|P|\sin(|X(g'g^{-1})|/\lo)\right)}{\sin(|X(g'g^{-1})|/\lo)} \pt
\ee
This expression clearly parallels that of (\ref{eq:angleprop}) coming from the spin representation, as we have anticipated to be the case in this simple example. In general, we remark again, we could expect such similarity only in the semi-classical limit, where we have the identification $|P| = 2(\hbar/\lo)j$.

On the other hand, we may plug the finite time propagator (\ref{eq:angleprop}) in terms of the spin label $j$ into the path integral to give the infinitesimal propagators. In this case, we obtain a first order path integral of the form
\bea
	\langle g',t'| g,t \rangle &=& \left[ \prod_{k=1}^{N-1} \int_{SO(3)}\!\!\pi^{2}\lo^{3}\dd g_{k} \right] \left[ \prod_{k=0}^{N-1} \sum_{j_{k}} \frac{1}{\pi^{2}\lo^{3}} \right] \nonumber\\
	&& \times \left[ \prod_{k=0}^{N-1} (2j_{k}+1) \frac{\sin((2j_{k}+1)|X(g_{k+1}g_{k}^{-1})|/\lo)}{\sin(|X(g_{k+1}g_{k}^{-1})|/\lo)} e^{-\frac{i}{\hbar}\epsilon \frac{2\hbar^{2}}{m\lo^{2}} j_{k}(j_{k}+1)} \right] \pt
\eea
Here we may approximate $|X(g_{k+1}g_{k}^{-1})|/\lo \approx \sin(|X(g_{k+1}g_{k}^{-1})|/\lo)$, and thus further write for the factors in the integrand
\be
	\frac{\sin((2j_{k}+1)\sin(|X(g_{k+1}g_{k}^{-1})|/\lo))}{\sin(|X(g_{k+1}g_{k}^{-1})|/\lo)} = \frac{2j_{k}+1}{4\pi} \int_{\mathbb{S}^{2}} \dd^{2}\vec{n}_{k}\ e^{i(2j_{k}+1)\vec{n}_{k} \cdot Y(g_{k+1}g_{k}^{-1})/\lo} \pt
\ee
Substituting back to the path integral, we obtain
\bea
	\langle g',t'| g,t \rangle &=& \lim_{N \rightarrow \infty} \left[ \prod_{k=1}^{N-1} \int_{SO(3)}\!\!\pi^{2}\lo^{3}\dd g_{k} \right] \left[ \prod_{k=0}^{N-1} \sum_{j_{k}} \frac{2\hbar/\lo}{(2\pi\hbar)^{3}} \int_{\mathbb{S}^{2}} \frac{\hbar^{2}}{\lo^{2}}(2j_{k}+1)^{2}\dd^{2}\vec{n}_{k} \right] \nonumber\\
	&& \quad \times \exp\left\{ \frac{i}{\hbar} \sum_{k=0}^{N-1} \epsilon \left[ \frac{\hbar}{\lo} (2j_{k}+1)\vec{n}_{k} \cdot \frac{Y(g_{k+1}g_{k}^{-1})}{\epsilon} - H(j_{k})\right] \right\} \pc
\eea
where $H(j_{k}) = (2\hbar^{2}/m\lo^{2})j_{k}(j_{k}+1)$. Again, this expression is analogous to the discrete first order path integral (\ref{eq:discpi}) we derived previously, with the identification $|P| = 2(\hbar/\lo)j$ (which would hold also in the general case, in the semi-classical limit). 

However, we notice that taking the continuum limit $N \rightarrow \infty$, $\epsilon \rightarrow 0$ and applying the calculus of variations to this (pseudo) path integral expression, in order to study the semi-classical limit is more difficult than in the true first order (non-commutative) path integral in terms of the variables $P$, since the spin labels $j_{k}$ take only integer values. It is also interesting to note that we obtain the standard Euclidean path integral from the previous expression by taking the Euclidean limit $\lo \rightarrow \infty$, $j_{k} \rightarrow \infty$, $j_{k}/\lo = const.$ with the identification $P_{k} = 2(\hbar/\lo)j_{k}$. This limit is both conceptually and mathematically different with respect to the semi-classical limit.

\

Next we will calculate the second order form of the path integral for the free particle in terms of the group (configuration) variables. For the Hamiltonian $\hat{H} = \hat{P}^{2}/2m$ the corresponding quantum corrected Hamiltonian is found to be $H_{q}(P) = (P^{2} + \frac{\hbar^{2}}{\lo^{2}})/2m$, which agrees with the expression in \cite{ChaichianDemichev} (when we set $\lo = 2$ to use the same length scale for the group as in \cite{ChaichianDemichev}) originally found by DeWitt \cite{DeWitt} in the case of a general curved space. Then, each of the $N$ integrals over $P_{k}$ in (\ref{eq:gdiscpi})
\bea
	\langle g',t' | g,t \rangle &=& \lim_{N \rightarrow \infty} \left[ \prod_{k=1}^{N-1} \int_{SO(3)}\pi^{2}\lo^{3}\dd g_{k} \right] \left[ \prod_{k=0}^{N-1} \int_{\mathbb{R}_{\star}^{3}} \frac{\dd^{3} P_{k}}{(2\pi\hbar)^{3}} \right] \nonumber\\
	& & \qquad \times \exp\left\{ \frac{i}{\hbar} \sum_{k=0}^{N-1} \epsilon \left[ \frac{Y(g_{k}^{-1}g_{k+1})}{\epsilon} \cdot P_{k} - \frac{1}{2m} \left(P_{k}^{2} + \frac{\hbar^{2}}{\lo^{2}}\right) \right] \right\}
\eea
becomes the usual Fourier transform of the function $\exp\left[-\frac{i\epsilon}{2m\hbar} \left(P_{k}^{2} + \frac{\hbar^{2}}{\lo^{2}}\right)\right]$ to the coordinate variables $Y(g_{k}^{-1}g_{k+1})$. Thus, by performing the Gaussian integrals we obtain
\bea
	\langle g',t' | g,t \rangle &=& \lim_{N \rightarrow \infty} \left[ \prod_{k=1}^{N-1} \int_{SO(3)}\!\!\pi^{2}\lo^{3}\dd g_{k} \right] \left[ \prod_{k=1}^{N} \left( \frac{m}{2\pi i\hbar\epsilon}\right)^{\frac{3}{2}} \exp\left\{ \frac{i\epsilon}{\hbar} \frac{m}{2} \left(\frac{Y(g_{k}^{-1}g_{k+1})}{\epsilon} \right)^{2} \right\} \right] e^{-\frac{i(t'-t)\hbar}{2m\lo^{2}}} \nonumber\\
	&=& \lim_{N \rightarrow \infty} \left[ \left( \frac{m}{2\pi i\hbar\epsilon}\right)^{\frac{3}{2}} \prod_{k=1}^{N-1} \int_{SO(3)}\frac{\pi^{2}\lo^{3}\dd g_{k}}{(2\pi i\hbar\epsilon / m)^{\frac{3}{2}}} \right] \exp\left\{ \frac{i}{\hbar} \sum_{k=1}^{N} \epsilon \frac{m}{2} \left(\frac{Y(g_{k}^{-1}g_{k+1})}{\epsilon} \right)^{2} \right\} e^{-\frac{i(t'-t)\hbar}{2m\lo^{2}}} \pt \nonumber\\
\eea
The product of integrals including the factors of $\left(\frac{m}{2\pi i\hbar\epsilon}\right)^{\frac{3}{2}}$ becomes the second order path integral measure at the continuum limit, as in the usual case of $\mathbb{R}^{3}$, and the function in the exponent becomes $i/\hbar$ times the classical action \cite{Schulman,Dowker}, since by defining $\bar{V}_{\epsilon}(k\epsilon)$ by $g_{k+1} \equiv \exp[i\epsilon\bar{V}_{\epsilon}(k\epsilon) \cdot \bar{\sigma}/\lo]\, g_{k}$, where $|\epsilon\bar{V}_{\epsilon}(k\epsilon)| < \frac{\pi}{2}\lo$, we have
\be
	\frac{m}{2}\left( \frac{Y(g_{k}^{-1}g_{k+1})}{\epsilon} \right)^{2} = \frac{m}{2} \bar{V}_{\epsilon}^{2}(k\epsilon) \stackrel{N \rightarrow \infty}{\rightarrow}  \frac{m}{2} \bar{V}_{0}^{2}(t) = -\frac{m}{2} \frac{\lo^{2}}{2}\tr\left( g^{-1}(t) \frac{\dd g}{\dd t}(t) g^{-1}(t) \frac{\dd g}{\dd t}(t) \right) = \mathcal{L}_{class}\pc
\ee
which is the classical Lagrangean of a free point particle on $SO(3)$. Finally, we can write for the continuum path integral
\be\label{eq:freepi}
	\langle g',t' | g,t \rangle = \int_{\substack{ g(t) \equiv g \\ g(t') \equiv g' }} \mathcal{D}g(t)\ \exp\left[ \frac{i}{\hbar} \int_{t}^{t'}\!\! \dd t\ \frac{m\lo^{2}}{2} \tr\big(\dot{g}^{2}(t)\big) -\frac{i(t'-t)\hbar}{2m\lo^{2}} \right] \pc
\ee
where $\mathcal{D}g(t)$ is the continuum limit of the path integral measure given above, and $\dot{g} := -ig^{-1} \frac{\dd g}{\dd t} \in \mathfrak{so}(3)$ is the velocity of the particle. This coincides with the path integral for free particle on $SO(3)$ obtained by other methods \cite{MarinovTerentyev,ChaichianDemichev}.

\section{Conclusions}\label{sec:conclusions}
We have shown, starting from the canonical formulation of quantum mechanics on $SO(3)$, that a first order path integral can be derived for a quantum system with configuration space $SO(3)$, using the non-commutative dual space variables via the group Fourier transform, which produces the correct semi-classical behavior, and is consistent with earlier results in the literature. 

\

The advantages of the approach we have studied here are the following: On one hand, it provides an alternative to the use of representation theory, and a more intuitive picture of the classical dynamics behind the quantum one with continuous non-commutative momentum variables, which resemble more closely the classical momentum variables. On the other hand, this approach makes the semi-classical analysis much more straightforward.

\

Our main interest in this formulation of the quantum dynamics of this simple system is in the fact that it exemplifies, as we have discussed, the respective role and use of the standard spin network/spin foam representation of canonical Loop Quantum Gravity and Spin Foam models, and the recent flux/non-commutative simplicial path integral representation of these theories. 

More precisely, what we think we can learn from the above analysis for quantum gravity models, Loop Quantum Gravity and Spin Foams in particular, is the following:

\begin{itemize}
\item Given the classical phase space of the theory, the use of the non-commutative Fourier transform and of the $\star$-product described above is very natural, and seems to provide a correct conjugate representation for quantum states.

\item  Similarly, it is not surprising, although certainly welcome, that the dynamics of the theory in terms of the dual non-commutative variables takes the form of the expected first order path integral. In the spin foam case, this is to be expected every time the quantum spin foam dynamics is supposed to encode the classical geometry of simplicial gravity at the covariant level (as it is the case for all current Spin Foam models).

\item In the case considered, the non-trivial phase space structure gives naturally rise to quantum corrections into the classical action in the path integral formulation of a quantum theory. Similar corrections are to be expected in the LQG/Spin Foam context. Moreover, our results support the choice of the specific $\star$-product structure we use, and thus of the associated (Duflo) quantization map (operator ordering) \cite{FreidelMajid,HannoThomas}, also in Loop Quantum Gravity. This may lead in general to different spectra for observables with respect to the commonly used ones. This point is also discussed, for example, in \cite{Alekseev}.

\item The use of non-commutative dual variables is going to be advantageous in the study of the semi-classical approximation of the theory, exactly because it brings the quantum dynamics in the form of a path integral. As we have seen, this also clarifies the reason behind the expression which spin foam amplitudes take in this approximation.

\item Finally, our results indicate that an additional parameter encoding the scale (volume) of the group manifold, here called $\lo$, which can be made explicit also in the Spin Foam/LQG context, could play an important role in studying further the commutative or continuum approximations of the theory, independent from the semi-classical one. 
\end{itemize}

\

All the above points deserve to be further explored in the quantum gravity context, on the basis of the intuition provided by our results. Further possible developments include extending the group Fourier transform to general Lie groups, in which case the avoidance of representation theory can become even more advantageous. Altogether, our results further indicate that the new non-commutative variables make sense, both mathematically and physically, and that the non-commutative methods can be applied successfully, in general, where found advantageous.

\section*{Acknowledgements}
We thank Carlos Guedes for useful discussions on cotangent bundles, $\star$-products and Fourier transforms. This work was supported by the A. von Humboldt Stiftung, through a Sofja Kovalevskaja Prize, which is gratefully acknowledged.

\end{document}